# First-principles-based Prediction of Phase Fields: Part I. Binary and Ternary Refractory Alloys

Pravan Omprakash[1], Nicholas Crnkovich[2], John Cavin[3], Nathan Curtis[2], Adrien Couet[2], Rohan Mishra[3,1]

[1]Institute of Materials Science & Engineering, Washington University in St. Louis, St. Louis, MO, USA

[2]Department of Nuclear Engineering & Engineering Physics, University of Wisconsin, Madison, WI, USA

[3]Department of Mechanical Engineering & Materials Science, Washington University in St. Louis, St. Louis, MO, USA

Corresponding authors: P.O. (o.pravan@wustl.edu); R.M. (rmishra@wustl.edu)

## Abstract

Multiple principal element alloys (MPEAs) exhibit complex phase equilibria involving multinary solid solutions and intermetallics, which makes it challenging to predict their temperature-composition phase diagrams. Their vast compositional space makes first-principles methods prohibitively expensive, while CALPHAD is limited by scarce experimental data. Here, we present a computationally efficient framework to predict the solvus phase boundaries, and hence, phase fields, in refractory MPEAs composed of Cr, Hf, Mo, Nb, Ta, Ti, V, W, and Zr. The approach combines DFT calculated binary mixing enthalpies with sub-regular solution models to construct phase diagrams without fitting higher-order interactions, enabling efficient scaling across composition space. Validation against 36 binary and 15 ternary phase diagrams demonstrates good agreement, with both experimental results and CALPHAD calculations. We find that the prediction accuracy is enhanced by incorporating lattice dependent energetics through sub-regular solution models and including temperature dependent elemental phase transitions. The framework captures miscibility gaps, solid-solution stability, and intermetallic formation, with predicted miscible temperatures typically within 300 K of experimental values. Overall, this work establishes a scalable, first-principles-based route for high-throughput prediction of phase diagrams in refractory MPEAs. A publicly accessible web interface has also been developed to allow interactive exploration of the predicted phase diagrams, available at https://raptor.engr.wustl.edu.

## 1. Introduction

Multiple principal element alloys (MPEAs) comprising of four or more elements mixed at near-equimolar concentrations have expanded the alloy-design space by opening up the unexplored central regions of multinary phase spaces [2]. These alloys often exhibit diverse microstructures containing both multinary solid solutions and ordered precipitates, and with appropriate design, can yield properties superior to those of conventional alloys [3-6]. Predicting the phases that can coexist as a function of the multinary composition and temperature is therefore critical, as phase equilibria directly govern material properties. In this context, temperature–composition phase diagrams can accelerate the development and optimization of MPEAs for targeted applications. However, experimental exploration of the vast compositional space of MPEAs is impractical [7], making reliable and computationally efficient modeling approaches essential. Calculation of Phase Diagrams, or CALPHAD, has traditionally been employed to predict alloy phase diagrams by fitting phenomenological models to empirical data, but, its applicability to MPEAs is constrained by the scarcity of experimental inputs [8]. Increasingly, first-principles-based computational approaches with information regarding solid-solution stability in MPEAs are being used to augment experimental thermodynamic databases [9-11]. However, these methods are limited in their ability to scale with compositional complexity, presenting challenges for high-throughput screening of MPEAs.

Prior studies [12-14] have shown that binary mixing enthalpies calculated using density-functional theory (DFT), combined with thermodynamic solution models, provide an efficient route to predicting solid-solution stability in *equimolar* MPEAs. The use of special quasirandom structures (SQS) to model chemical disorder [15] allows accurate DFT calculations of mixing enthalpies in modest-sized supercells, and the construction of thermodynamic databases. Zhang et al. [12] employed regular solution models to describe the mixing enthalpy of refractory BCC solid solutions, and applied this framework to predict the phase stability of ~20,000 equimolar MPEAs, including $NbVZrTi_{(1-x)}$ alloys whose phase evolution was subsequently validated experimentally. Chen et al. [13] applied a similar approach to identify 30,201 potential single-phase equimolar quinary alloys. Their model achieved an accuracy of 74% for predicting the solid solution stability of >100 experimentally reported equimolar alloys. However, whether such binary regular solution models can be extended to predict full temperature–composition phase diagrams beyond the equimolar limit, and identify phase fields efficiently, remains an open question. Addressing this challenge requires a computationally tractable yet robust framework capable of predicting complete phase diagrams, thereby enabling high-throughput MPEA design and discovery. This capability is most consequential in the non-equimolar regime, where tuning constituent concentrations away from equimolarity enables optimization of phase fractions, solid-solution stability, and properties, thereby, unlocking the full potential of MPEAs.

In this Article, we present a first-principles-based framework capable of predicting the solvus phase boundaries across compositional spaces and temperatures for refractory binary and ternary alloys containing any combination of the nine elements typically used in refractory MPEAs: Cr, Hf, Mo, Nb, Ta, Ti, V, W, and Zr. We combine DFT-calculated binary mixing enthalpies at various compositions with a sub-regular solution model to predict the 36 binary and 86 ternary phase diagrams, and benchmark them against experimental phase diagrams, where available, or else, with PANDAT-based phase diagrams [16]. We demonstrate that convex-hull approaches employed in previous studies to predict equimolar MPEAs [12-14] require modifications to accurately capture phase diagrams. Specifically, we find that regular solution models are sufficient to describe binary mixing enthalpies only when both the elements share the same stable lattice; when they don't, sub-regular solution models with two fitting parameters are necessary. We also show that temperature-dependent phase transitions of elements become important to describe the stability of their alloys with temperature, and that this can be easily incorporated by changing the reference energy of the elements at the appropriate transition temperatures from experimental databases. With these corrections, our model reliably captures miscibility gaps, solid-solution stability, and intermetallic formation, with the majority of predicted miscibility temperatures lying within ±300 K of experimental values. We attribute the few systems where the predicted phase fields deviate from experiments primarily to phase transitions of their intermetallics with temperature. Using the case of $ZrV_2$, which undergoes a transition from a rhombohedral to cubic Laves phase at ~100 K, we show that these can be readily incorporated in our framework. In Part II, we extend the framework to higher-order MPEAs and conduct a meta-analysis across thousands of predicted phase fields to identify systematic phase-stability trends. Together, these developments demonstrate scalability of the framework and its applicability to high-throughput phase-diagram prediction for refractory MPEAs, to eventually help accelerate the design of functional alloys with desired microstructures. Additionally, a publicly accessible web-based interface has been developed to facilitate interactive exploration of the predicted phase diagrams, accessible at https://raptor.engr.wustl.edu .

## 2. Methods

### 2.1. Model for predicting the Gibbs free energy of multinary solid solutions

The Gibbs free energy of solid solutions, $G(x,T)$, can be minimally expressed as a sum of three primary components that are themselves a function of the mole fraction, $x$, and/or the absolute temperature, $T$: 1) the mixing enthalpy, $\Delta H_{mix}(x)$, 2) the configurational entropy, $\Delta S_{mix}(x)$, and 3) temperature-dependent free energies of the different phases of the unary end members, $G_o^i(T)$, where $i$ represents an element. $G_o^i(T)$ for most unary endmembers have been obtained from fitting to experimental data and therefore contain enthalpic and vibrational entropic contributions. Here, we have ignored other entropic contributions, such as from vibrations, magnetism, and electrons, for two reasons. Firstly, the electronic and spin-contributions to entropy are typically two orders of magnitude smaller than the configurational entropy [17]. Vibrational energy on the other hand can have a sizeable contribution to the energy on a case-to-case basis [17, 18], however evaluating these energies for each solid solution/ordered compound composition requires expensive computational calculations, inhibiting their use for high-throughput screening. In subsequent sections, we discuss cases where vibrational energy affects the predicted solvus phase boundaries. Hence, the $G(x)$ for a hypothetical alloy, $A_{x_A}B_{x_B}C_{x_C}D_{x_D}$, can be written as shown in (1):

$$G(x,T) = \Delta H_{mix}^{ABCD}(x) - T\Delta S_{\text{mix}}^{ABCD}(x) + \sum_i G_o^i(T) \tag{1}$$

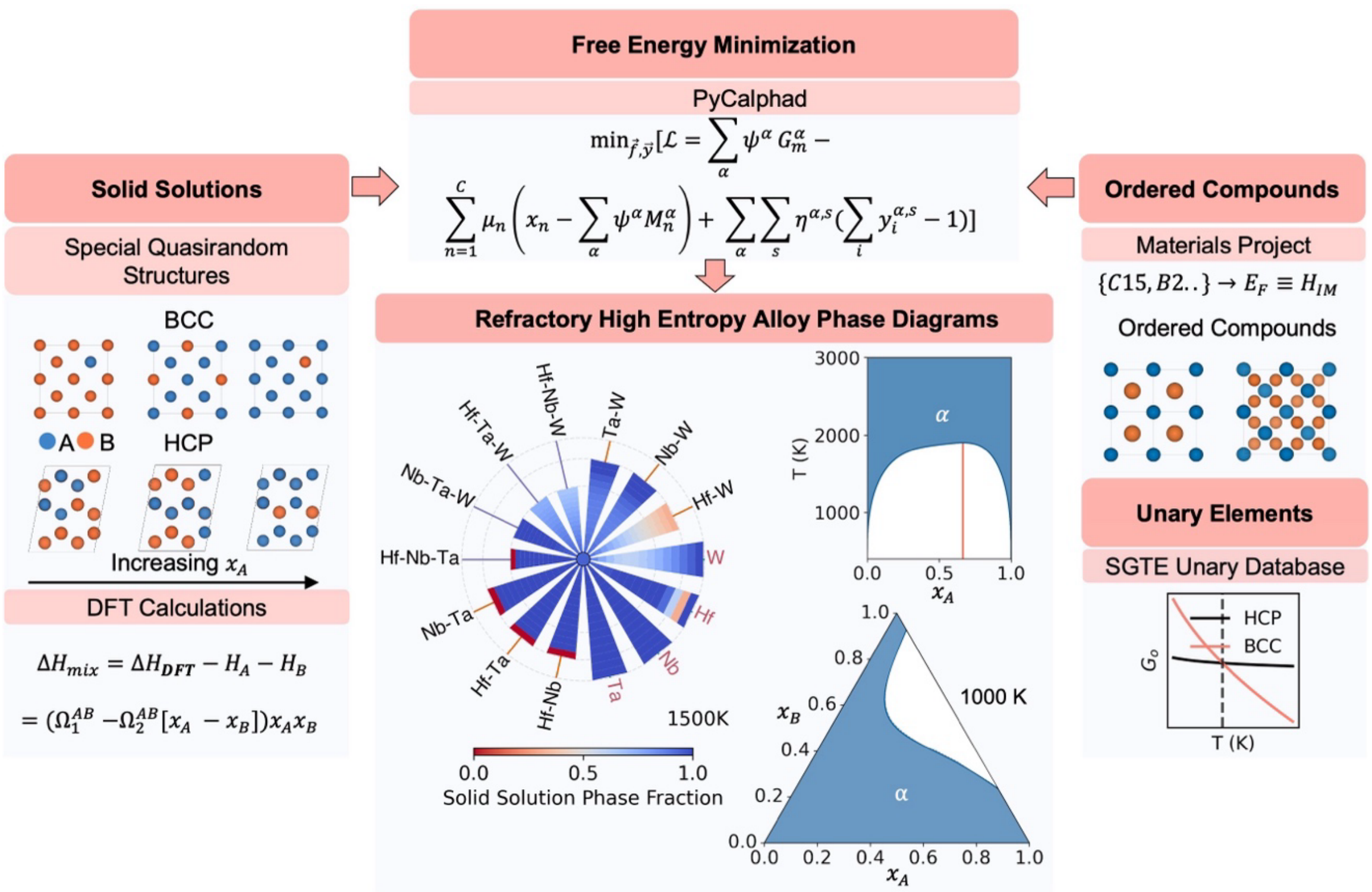

**Fig. 1. Schematic of the workflow used for the rapid prediction of phase diagrams of MPEAs.** The top-left panel shows disordered binary alloys simulated using special quasirandom structures (SQS) on competing lattices for BCC and HCP. The SQS approximates a random atomic arrangement at a chosen composition, providing a tractable supercell for DFT calculations. Then the DFT total energies are used to extract the composition-dependent mixing enthalpy, $\boldsymbol{\Delta H_{mix}}$, shown in the bottom-left panel. Competing ordered intermetallics are incorporated from the Materials Project database, as shown in the top-right panel. The bottom-right panel shows that the Gibbs free energies of elements are taken from the SGTE unary database, which provides consistent elemental thermodynamics and temperature-dependence for all the elementary phases. All phase models (solid solution phases, ordered compounds and elementary phases) are fed into *pycalphad*, where equilibrium is obtained by minimizing the total Gibbs free energy, $\boldsymbol{G(x, T)}$, to generate temperature-composition phase diagrams, as shown in the middle panel.

The configurational entropy as a function of *x* can be written as shown in (2):

$$\Delta S_{\text{mix}}^{ABCD} = -k_B \sum_i x_i ln(x_i) \tag{2}$$

We estimate the mixing enthalpy for an MPEA, $\Delta H_{mix}^{ABCD}(x)$, from the constituent binary mixing enthalpies, modeled with a sub-regular solution model, as shown in Fig. 1. Previously, we and others have used a regular solution model to calculate the phase stabilities of equimolar MPEAs [12-14]. Here, we extend this approach to predict phase diagrams for MPEAs at any composition using sub-regular solution models. We propose that binary sub-regular solution models are sufficient to describe the interactions of a multinary solid solution. Eqns. (3) and (4) describe the regular [19] and sub-regular solution models [20], respectively, for a binary solid solution, *AB*.

$$\Delta H_{mix}^{AB}(x_A) = \Omega^{AB} x_A(1 - x_A), \tag{3}$$

where, $\Omega^{AB}$ represents an interaction parameter.

$$\Delta H_{mix}^{AB}(x_A) = [(\Omega_0^{AB} + \Omega_1^{AB}(2x_A - 1)]x_A(1 - x_A), \tag{4}$$

where $\Omega_0^{AB}$ and $\Omega_1^{AB}$ are the binary interaction parameters as modeled by a sub-regular solution model. There are other forms for the sub-regular solution [21]—all of them represent a cubic fit—but we choose this form, because it is intuitive; as $\Omega_1^{AB} \to 0$ , the sub-regular solution collapses to the regular solution model. This intuition lends itself to multiple physical meanings. Cavin et al. [21] demonstrated that a sub-regular solution model fitted to DFT-calculated binary mixing enthalpies can rapidly predict binary phase diagrams of transition-metal dichalcogenides. The authors discussed that the cubic sub-regular model captured three-body effects that are important in covalently bonded systems. In this work, we put forward another physical meaning, that of composition-dependent interaction parameters for alloys having metallic bonding unlike directional bonds in covalent systems. In a regular solution model, the interaction parameter, $\Omega_0$, can be related to the difference between like and unlike bond energies as $\Omega_o = z(\epsilon_{AA} + \epsilon_{BB} - 2\epsilon_{AB})$, where $\epsilon$ is the bond energy between two similar (*AA, BB*) or different (*AB*) atoms, and *z* is the first-nearest neighbor coordination number. In the sub-regular model, both interaction parameters, $\Omega_0$ and $\Omega_1$, can hence be combined to form a composition-dependent interaction parameter, $\Omega(\mathrm{x}) = \Omega_{\mathrm{o}} + \Omega_1(2x_A - 1)$. This formulation implies that the interactions between *A–B* varies with the local chemical environment, or, equivalently, with alloy composition. Thus, the sub-regular interaction parameters encodes how the effective chemical interaction between unlike atoms changes as the surrounding alloy environment changes. The implication of this physical meaning will be further discussed in Section 4.1. Thus, by considering sub-regular solution models, we treat mixing enthalpies as continuous functions of mole fraction—curves in binary systems, sheets in higher-dimensional composition space—rather than specific values at a fixed mole fraction.

The interaction parameters for each binary solid solution can then be obtained from DFT-calculated mixing enthalpies. We use SQS models [15] to generate supercells that approximate disordered binary solid solutions on BCC and HCP lattices, at five different compositions, $x = [0.125, 0.25, 0.5, 0.75, 0.875]$, as depicted in the left panel of Fig. 1. We don't consider the FCC lattice, as the alloys formed from the pool of elements of interest do not form FCC phases [13]. Following previous studies [12-14], we use a 24-atom supercell for the BCC lattice that shows a perfect match to a random alloy for pairwise interactions up to the 5$^{\text{th}}$ nearest neighbor and triplet interactions up to the 3$^{\text{rd}}$ nearest neighbor. For the HCP lattice, we use a 16-atom supercell, considering pair interactions up to the 6$^{\text{th}}$ nearest neighbor, and triplet interactions up to the 3$^{\text{rd}}$ nearest neighbor. The SQS models are generated using the Alloy Theoretic Automated Toolkit (ATAT) package [22]. The computational details for performing the DFT calculations are given in Section 2.3.

The multinary mixing enthalpy, $\Delta H_{mix}^{ABCD}$, can then be estimated from the binary interaction parameters, as shown in Eqn. (5):

$$\Delta H_{mix}^{ABCD}(x) = \sum_{i,j\in\{ABCD\},i\neq j}\left[\Omega_1^{ij} + \Omega_2^{ij}(x_i - x_j)\right]x_i x_j. \tag{5}$$

Finally, the Gibbs free energies for the unary members, $G_o^i(T)$, are obtained from the elemental temperature-dependent enthalpies provided in Scientific Group Thermodata Europe (SGTE) database [23]. The SGTE unary database also provides temperature-dependent reference data for modeling elemental phase transitions. For our refractory systems, the most significant transition is from HCP to BCC in Ti, Zr, and Hf, as illustrated in the bottom right panel of Fig. 1.

## 2.2. Predicting phase diagrams of refractory multi-principal element alloys

The phase diagram of an alloy can be obtained by calculating *G(x)* for the constituent phases and then performing a multi-phase, multi-component minimization of the free energy as a function of composition and temperature. In this work, the minimization is carried out using the Python-based, open-source package *PyCalphad* [24]. The equation minimized is shown in Eqn. (*6*), and also in Fig. 1:

$$\min_{\vec{f},\vec{y}}\left[\mathcal{L} = \sum_{\alpha}\psi^{\alpha}G^{\alpha} - \sum_{n=1}^{C}\mu_n\left(x_n - \sum_{\alpha}\psi^{\alpha}M_n^{\alpha}\right) + \sum_{\alpha}\sum_{s}\eta^{\alpha,s}\left(\sum_{i}y_i^{\alpha,s} - 1\right)\right]. \tag{6}$$

The first term consists of the sum of free energies of all the listed phases, where $G^{\alpha}$ is the Gibbs free energy of a stable phase $\alpha$, and $\psi^{\alpha}$ is the number of formula units for the phase. The minimization of this Gibbs free energy term has to happen under a set of internal constraints, such as fixed composition, which is considered using Lagrange multipliers, as represented by the second and third terms in Eqn. (*6*). The second term is the constraint of mass balance, where $\mu_n$ is the chemical potential of component *n,* $x_n$ is the fraction of component *n*, $M_n^{\alpha}$ is the amount of *n* in phase $\alpha$ and *C* is the total number of components. Note that for this term the Lagrange multiplier is the chemical potential $\mu_n$. The third term is another constraint—the sum of constituent fractions in each sublattice of $\alpha$ should be unity—where *s* is the sub-lattice index, $\eta^{\alpha,s}$ is the corresponding Lagrange multiplier, and $y_i^{\alpha,s}$ is the constituent fraction of component *i* on sublattice *s* in $\alpha$. A more comprehensive description of this equation and the theoretical steps for minimizing it is discussed and implemented in the OpenCalphad framework [25].

*PyCalphad* requires a thermodynamic database (.TDB) file as input, which we generate using a Python script developed for this purpose. The Python script along with all the other related codes and raw data pertaining to the results presented in this work are available in a publicly available repository [26]. Additionally, a web interface is also made available to obtain predicted phase diagrams and related analysis for the alloys studied in this work at https://raptor.engr.wustl.edu . The TDB files contain binary interaction parameters for BCC and HCP solid solutions, along with the formation energies ($H_f$) of stable binary intermetallics obtained from the first-principles DFT calculations for the alloys considered in the pool of elements. For creating a larger dataset of alloy information, Materials Project database [27] contains a vast repository of binary and ternary intermetallics whose formation energies can be used. We do not include temperature-dependent free energies of intermetallics, and therefore only consider intermetallic phases that are stable at 0 K, with one exception: the $ZrV_2$ C15 Laves phase. For this system, the free energy was calculated at 300 K using DFT-based phonon calculations, since the phase is unstable at 0 K [28]. Elemental data for the TDB files were taken from the SGTE database.

We calculated binary interaction parameters for the nine elements commonly found in refractory alloys: Cr, Hf, Mo, Nb, Ta, Ti, V, W, and Zr. Since most of these alloys adopt the BCC structure at high temperatures, all 36 BCC binary pairs were fit with the sub-regular solution model. For HCP interactions, 21 pairs containing Ti, Hf, and Zr were also fit with the sub-regular model, as these elements are HCP at room temperature but transition to BCC at elevated temperatures [29]. The remaining 15 pairs on the HCP lattice (Cr-W, Cr-Mo, Cr-Nb, Cr-V, Cr-Ta , Mo-Nb, Mo-Ta, Mo-V, Mo-W, Nb-W, Nb-Ta, Nb-V, Ta-W, Ta-V, V-W), which do not form HCP phases at any temperature, were modeled using a simpler regular solution form, which was found to give a satisfactory fit to the DFT energies. These alloys do not stabilize in the HCP phase at any temperature or composition because both constituent elements stabilize in the BCC lattice, and have high, positive free energies for the HCP lattice. The HCP phase of these alloys is also dynamically unstable and readily transforms to BCC [30]. Thus, a single calculation at $x = 0.5$ is performed ensuring that the HCP symmetry is preserved by placing constraints on the shape of the unitcell.

To calculate the energies of multiple compositions of these mechanically unstable structures, special approaches have been proposed [31] that rely on multiple nudged elastic band and phonon calculations to first determine the unstable modes in the structure and then determine the free energy. These calculations are computationally expensive and thus have been omitted here to keep the process fast. Using the calculated binary parameters, we generated composition–temperature phase diagrams for 36 binary, 84 ternary, 126 quaternary, and 126 quinary systems, which were then compared with experimentally reported phase diagrams, where available. Liquidus curves were omitted, as high-throughput DFT modeling of liquids currently remains computationally infeasible [32].

### 2.3. Computational details

We performed the first-principles DFT calculations using the Vienna Ab initio Simulation Package (VASP) [33] and the generalized gradient approximation (GGA) as implemented in the Perdew-Burke-Ernzerhof (PBE) exchange-correlation functional [34]. We used a plane-wave energy cutoff of 520 eV and relaxed the ionic positions until the forces were less than 0.01 eV/Å. Electronic self-consistent field calculations were considered converged when the total energy difference between successive iterations was less than $10^{-6}$ eV. The Brillouin zone was sampled using a Γ-centred *k*-points mesh with a spacing of 0.025/Å. 21 out of the 36 systems containing Ti, Zr or Hf were found to be unstable in the HCP or BCC lattices at certain compositions, leading to a loss of the parent lattice symmetry and relaxation into highly distorted configurations [30]. To address this issue, additional constraints were imposed to fix the cell shape, while allowing the lattice parameters and atomic positions to relax. A few off-equimolar compositions for the HCP lattice—except in Ti-Zr, Ti-Hf, and Hf-Zr—consisting of elements that exist only in the BCC phase failed to converge. Even with the additional constraints on cell shape, the atoms deviated heavily from the lattice points and created highly distorted cells. These points were discarded, but all binary systems were ensured to have at least three compositions — that are needed to fit a sub-regular solution model. All the DFT calculations performed for compiling this database are provided in the publicly available repository accompanying this article [26].

After obtaining the DFT energies at different compositions, we calculate $\Delta H_{mix}^{AB}(x)$ with respect to the energy of the constituent elements existing in the same lattice, $E(A)$ and $E(B)$, as shown in Eqn. (7):

$$\Delta H_{mix}^{AB}(x) = E\left(A_{x_A}B_{x_B}\right) - x_A E(A) - x_B E(B). \tag{7}$$

The binary interaction parameters from the mixing enthalpies can be obtained using Eqn. (3) and (4). The fitting of the solution models were performed using the *curve_fit* function implemented in the NumPy and SciPy libraries.

Phonon calculations were performed on the C15 Laves phase in the V-Zr system to model its temperature-dependent properties. In this approach, the partition function, *Z*, is calculated from the phonon density of states, and the vibrational free energies, *F*, are obtained from statistical thermodynamics equations derived from *Z*, as shown in Eqn. (8) and Eqn. (9):

$$Z = \exp\left(-\frac{\eta}{k_{\mathrm{B}}T}\right)\prod_{\mathrm{i}}\left[\frac{\exp\left(-\frac{(\hbar\,\omega_{\mathrm{i}})}{(2\,k_{\mathrm{B}}\mathrm{T})}\right)}{\left(1 - \exp\left(-\frac{(\hbar\,\omega_{\mathrm{i}})}{(k_{\mathrm{B}}\mathrm{T})}\right)\right)}\right], \text{ and} \tag{8}$$

$$F = -k_{\mathrm{B}}\mathrm{T}ln(\mathrm{Z}), \tag{9}$$

where, $\eta$ is the phonon density of states, $k_{\mathrm{B}}$ is the Boltzmann constant, ħ is the reduced Planck constant, and $\omega$ is the zone-center frequency of phonon mode *i*. We used the finite-displacements method on a 96-atom supercell to calculate the phonon band structure. All calculations were spin-polarized and initialized with a ferromagnetic configuration.

## 3. Results and Discussions

### 3.1. Impact of temperature-driven elemental phase transitions on solid-solution stability

Zhang et al. [12], Bokas et al. [14], and Chen et al. [13] have employed convex-hull constructions consisting of unary elements, lower order equimolar solid solutions, and binary and ternary intermetallics to assess the phase stability of multinary alloys. While this approach is fast and reasonably predictive for equimolar multinary alloys, it tends to underestimate the miscible temperature, $T_{misc}$, which is defined as the minimum temperature above which a single phase solid solution becomes stable for a particular composition. In Appendix A, we show that using this simplified convex hull approach can lead to a change in $T_{misc}$ of up to 700 K for quinary alloys, necessitating denser grids. However, convex-hull approaches do not scale linearly with addition of components or increasing the number of grid points. Therefore, we use the *Pycalphad* package [24] that efficiently minimizes the multiphase, multicomponent Gibbs free energies, as discussed in Section 2.2.

Next, we address the importance of modeling temperature-dependent phase transitions of the elements, particularly the HCP-to-BCC transition observed in Ti, Hf, and Zr at higher temperatures. The BCC phases of these elements are mechanically unstable at 0 K, but become stable at temperatures above 1000 K due to anharmonic phonons [29]. Incorporating these temperature-dependent reference energies, shifts the $G(x)$ curves—represented as sheets in multinary systems—along the temperature axis. The Ti-Ta binary alloy phase diagram is provided in Fig. 2 to illustrate the effect of the HCP→BCC transition of Ti at 1073 K. In the uppermost panel of Fig. 2(a), the Gibbs free energy of the BCC and HCP solid solutions of Ti-Ta at 0 K are plotted in pink and green, respectively. The dashed lines show the ideal mixing between the two elements in the two crystal structures. The HCP curve shows a negative deviation, and the BCC curve exhibits a positive deviation at 0 K. If the HCP to BCC transition of Ti is not included, then at finite temperatures, the only temperature dependence of the Gibbs free energy comes from the configurational entropy, as illustrated in the middle panel of Fig. 2(a) for 300 K. Entropy flips the positive deviation of the free energy of the BCC solution at 0 K to negative deviation at 300 K, and also reduces the free energy of the HCP solution. However, the alloy shall still segregate into a Ti-rich phase with HCP lattice and a Ta-rich phase with BCC lattice at 300 K, as seen in the resulting phase diagram in the upper panel of Fig. 2(b). In fact, if we ignore the Ti phase transition, then equimolar $Ti_{0.5}Ta_{0.5}$ will never stabilize as a single phase solid solution at any temperature—contradicting experimental observations [35]. Incorporating the phase transition of Ti, reduces the relative stability of the HCP phase, resulting in the stabilization of BCC Ti at 1073 K. In the lower panel of Fig. 2(a), the free energy curves at 300 K are plotted after considering the phase transition of Ti. The black arrow indicates the lowering of energy of BCC Ti. The resulting phase diagram shown in the lower panel of Fig. 2(b) is dramatically different from that in the upper panel, but is in excellent agreement with the experimental phase diagram [1].

We show the differences in miscible temperatures, $\Delta T_{misc}$, with and without considering the unary phase transitions, in Fig 2(c) for binary to quinary alloys. The median of $\Delta T_{misc}$ decreases from 0 K for binaries to –300 K for quinaries, as seen in Fig. 2(c). The negative sign implies that including the transitions, on an average, lowers the $T_{misc}$. Thus, inclusion of the temperature-dependent phase transitions for Hf, Ti, and Zr, can stabilize the BCC solid solution, and thereby enlarge the solid solution phase fields, at much lower temperatures. To model these phase transitions, we use the elemental data in the SGTE database that are derived from experimental enthalpy data [23]. While we explore refractory systems within the scope of this work, the use of temperature-dependent phase transitions in predicting solid solution phase fields is also important for elements like Fe and Mn, that show multiple FCC to BCC phase transitions at higher temperatures.

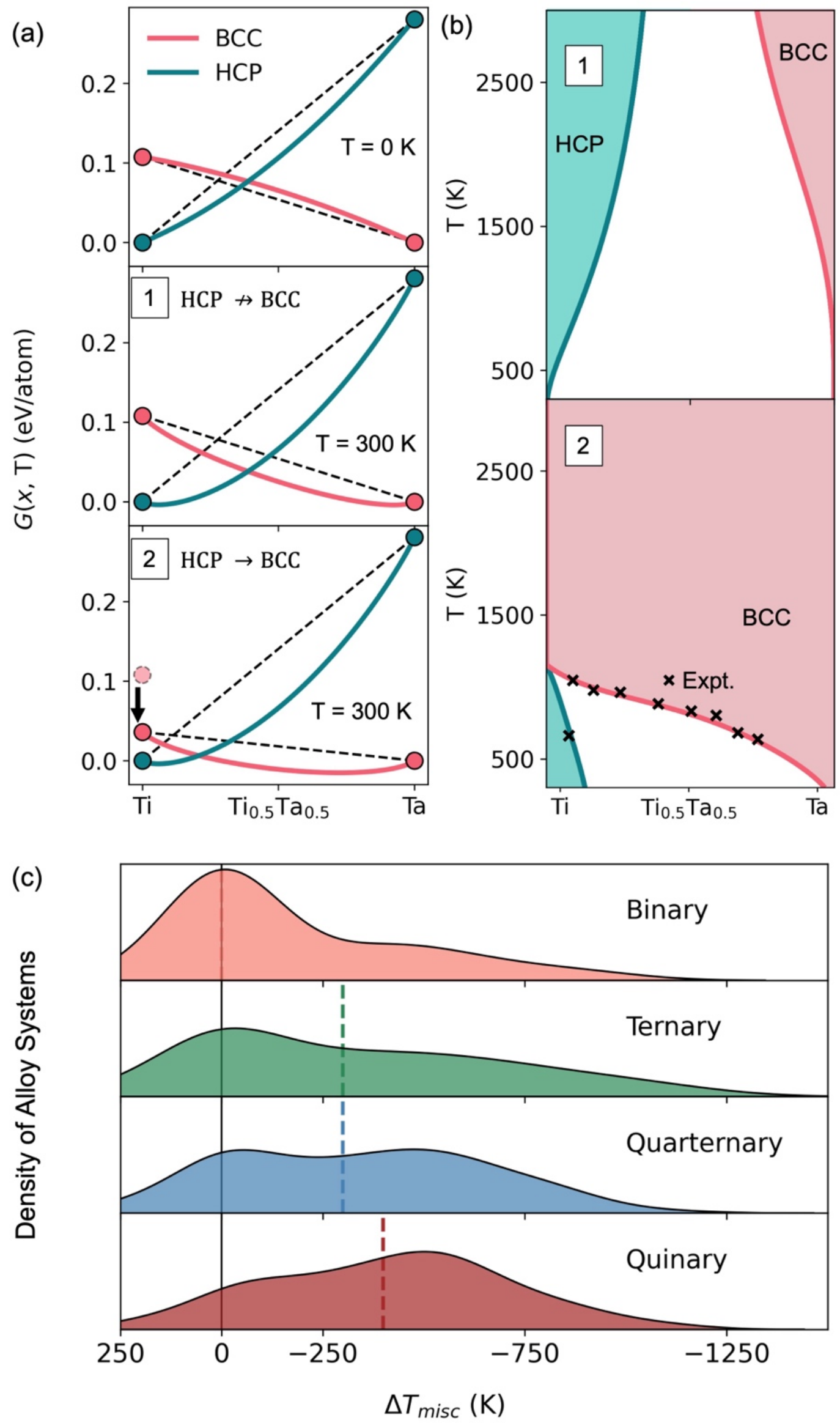

**Fig. 2. Influence of temperature-dependent elemental phase transitions on phase stability in alloys.** a) Gibbs free energy of the HCP (green) and BCC (pink) phases for the Ti-Ta binary system: top panel at 0 K, middle panel at 300 K without accounting for the HCP→BCC transition of Ti, and bottom panel at 300 K with the transition included. The black arrow indicates the reduction in the energy of the Ti BCC phase. Dashed black lines indicate ideal mixing of the two elements, and the colored lines show deviations from ideal mixing. b) Corresponding changes in the phase diagram after incorporating the phase transition. The shaded regions indicate single-phase solid solutions, and the white regions indicate phase separation. Experimental phase boundaries are marked in the phase diagram that includes the HCP→BCC transition of Ti obtained from Ref. [1] c) Distribution of the change in miscible temperatures, for binary, ternary, quaternary, and quinary equimolar compositions. Negative values indicate that the miscible temperature reduced after including the temperature-dependent unary phase transitions. The more negative $\Delta T_{misc}$ is, the higher the stability of the solid solutions. The densities of the alloy systems are normalized between 0 and 1.

### 3.2. Sub-regular solution model to capture the asymmetry in $\Delta H_{mix}$ with composition

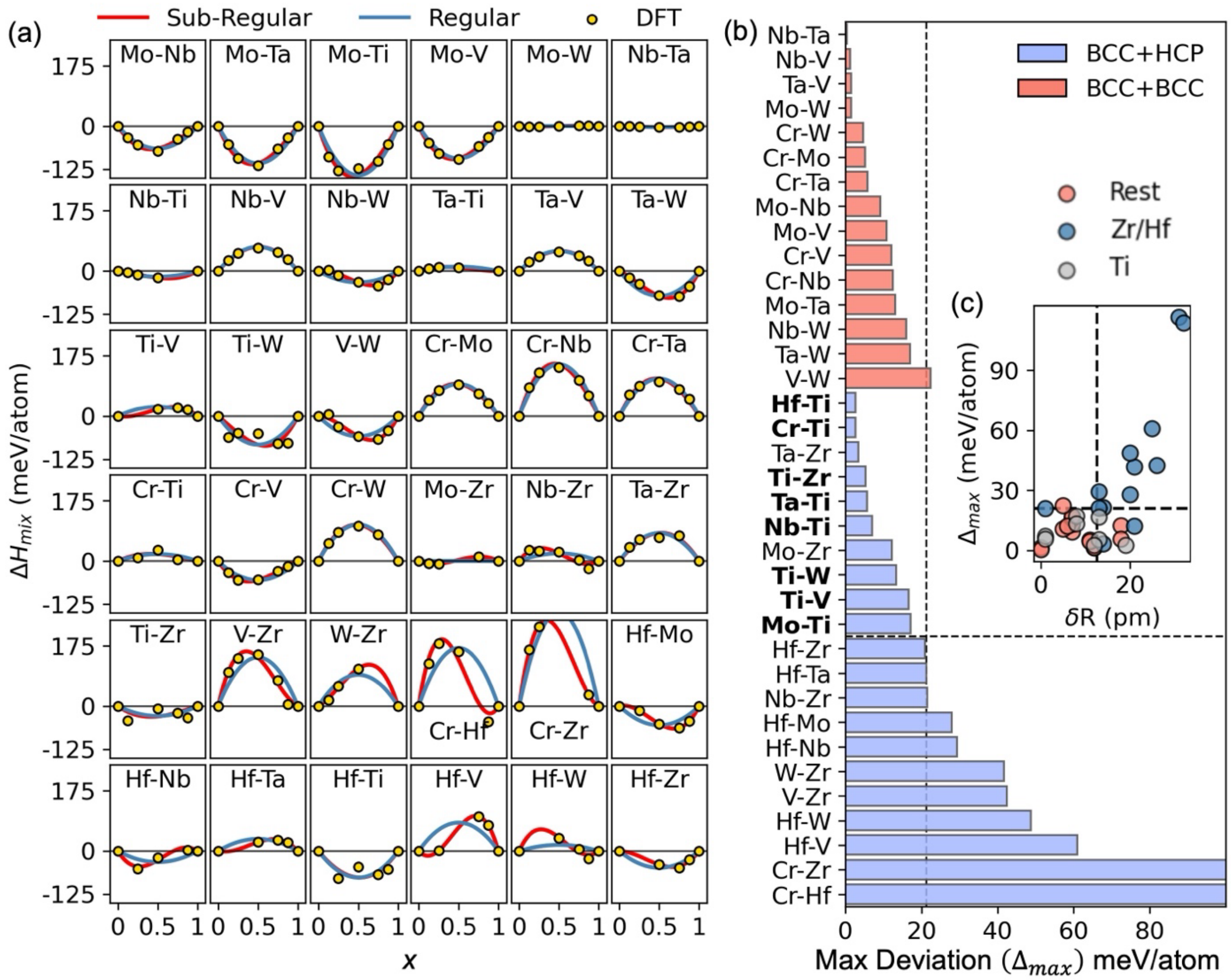

**Fig. 3. Assessment of regular vs sub-regular thermodynamic behavior in binary alloys.** (a) Composition-dependent mixing enthalpy $\Delta H_{\text{mix}}$ for the BCC lattices of the 36 binary systems. (b) Histogram of the maximum deviation ($\Delta_{max}$), defined as the maximum absolute difference in energies between the regular and sub-regular solution models. Blue bars (BCC+HCP) correspond to systems where the end members prefer different lattices, while red bars indicate BCC+BCC systems. Pairs containing Ti are highlighted in bold. The dashed, vertical line indicates the average $\Delta_{max}$ for all 36 pairs, while the dashed, horizontal line demarcates all systems that fall over this average value. (c) Correlation between the maximum deviation and the difference in metallic radii ($\delta R$) between the two constituent elements in each binary alloy. Highlighted systems involving Zr/Hf (blue) and Ti (grey) emphasize the role of atomic size mismatch in driving deviations from regular-solution behavior.

Having established the importance of modeling the temperature-dependent HCP-BCC transition of the relevant elements, we now turn to binary solid solutions. The regular solution model, given in Eq. (3), is a parabolic fit that produces a symmetric curve with the maximum deviation from ideal mixing at $x = 0.5$. In contrast, the sub-regular solution model is a cubic fit that can capture the asymmetry in $\Delta H_{mix}(x)$ [21]. To obtain the interaction parameters for binary solid solutions, we fit both regular and sub-regular solution models to the 36 binary pairs in the BCC lattice and 21 binary pairs in the HCP lattice. All the fits are shown in Fig. 3(a) and Fig. 9 in Appendix B for BCC and HCP lattices, respectively. The sub-regular solution model can describe the $\Delta H_{mix}(x)$ for nearly all the systems. A notable exception is the Ti–W BCC lattice, which shows significant asymmetry. Ångqvist et al. [36] provided a detailed account of the thermodynamic energies of this system obtained from a cluster expansion analysis, also reporting strong asymmetry and a large, negative BCC mixing enthalpy. The reason for this strong asymmetry in Ti-W requires further analysis. For such cases, a fourth-order fit requiring three interaction parameters may be more appropriate.

To assess the need for a sub-regular solution model, we examine the degree of asymmetry in the binary $\Delta H_{mix}(x)$ curves. Three scenarios can occur when fitting regular and sub-regular solution models to the DFT-calculated energies. In the first scenario, small concentrations of element *A* ($x < 0.5$) segregate from element *B* as a result of slightly positive mixing enthalpies. At higher concentrations of *A* (or equivalently, lower concentrations of *B* in an *A*-rich lattice), segregation is stronger due to large positive values of $\Delta H_{mix}(x)$. Such asymmetry is observed in several systems, including W-Zr, Hf-V, Cr-Zr, V-Zr, Hf-Ta, and Hf-W, as shown in Fig. 3(a). In the second case, the asymmetry is even more pronounced, with *B*-rich compositions exhibiting positive $\Delta H_{mix}(x)$ and *A*-rich alloys showing negative values, as seen in the HCP $\Delta H_{mix}(x)$ of Ta-Zr, Cr-Hf, Hf-W and Mo-Zr in Fig. 10. Finally, a third scenario is possible when the regular and sub-regular fits agree closely, such that both *A*-rich and *B*-rich compositions yield similar $\Delta H_{mix}(x)$. Such a situation occurs in almost all plots from Mo-Nb to Ta-Zr in Fig. 3(a). As discussed in Section 2, the sub-regular solution model may be interpreted as a composition-dependent regular solution model, where *A*-rich and *B*-rich lattices interact differently with their counterparts.

We now proceed to quantify the degree of asymmetry. As a quantitative metric, we use the maximum deviation in the mixing enthalpy ($\Delta_{max}$) between the regular and sub-regular fits: $\Delta_{max} =$

$|\max(\Delta H_{subreg}(x) - \Delta H_{reg}(x)|$, for each of the 36 binary pairs in the BCC lattice. The results are summarized in Fig. 3(b). In this plot, binary pairs containing elements that are stable in the BCC lattice at 0 K are shown in red, while pairs containing Ti, Hf, or Zr, i.e., BCC+HCP pairs, are shown in blue. The average $\Delta_{max}$ across all systems is 21 meV/atom, which corresponds to a thermal energy of ~250 K. This is within the ±300 K deviation between the predicted and experimentally measured miscibility temperatures discussed later. Notably, all the red points (BCC+BCC pairs) lie below this average. An exception is the V–W pair, which exhibits a large negative $\Delta H_{mix}(x)$ and correspondingly a larger deviation. This asymmetry does not affect the phase diagram in any significant way, as V-W remains miscible throughout the composition space.

The energetic deviations shown in Fig. 3 quantify how strongly the DFT-calculated mixing enthalpies depart from the symmetric form imposed by the regular solution model. However, energetic deviations do not directly reveal their impact on the predicted phase diagram. We therefore compared the miscibility temperatures obtained from regular and sub-regular fits at $x = [0.25, 0.50, 0.75]$ for each binary system. The miscible temperature, $T_{misc}$, is the temperature at which an alloy stabilizes into a single-phase solid solution at a certain composition. We used the maximum absolute difference of miscible temperatures from the two models, maximum $|\Delta T_{misc}|$, as a measure of sensitivity of the phase boundaries to the solution models.

As shown in Fig. 9(a), $\Delta_{max}$ correlates linearly with the resulting change in miscibility temperature, with a Pearson correlation of 0.79. The system-resolved distribution of maximum $|\Delta T_{misc}|$ in Fig. 9(b) further shows that most BCC+BCC binaries have comparatively small sensitivity to the choice of the solution model, with an average change of 253 K—with 14 out of 15 systems showing a maximum change under 500 K—whereas 11 out of 21 BCC+HCP systems exhibit shifts greater than 500 K. Thus, the asymmetry in the mixing enthalpy is a viable measure of the corresponding change in the phase field boundaries.

11 blue pairs, particularly those containing Hf and Zr, exhibit high $|\Delta T_{misc}|$ and high-$\Delta_{max}$, indicating that systems with elements preferring different lattices require a sub-regular model to capture their mixing enthalpies. This observation is physically intuitive: introducing large amounts of an HCP-forming element destabilizes the BCC lattice, resulting in larger values of $\Delta H_{mix}(x)$ at higher $x$, as compared to lower $x$ values. Interestingly, pairs containing Ti do not show pronounced asymmetry and correspondingly lower values of maximum $|\Delta T_{misc}|$. The origin of this behavior is not immediately clear. One possible explanation could be the difference in atomic radii between the elements. A larger size mismatch may correlate with a higher degree of asymmetry, since smaller atoms can more easily fit into the lattice of primarily larger atoms, whereas the reverse may be unfavorable, leading to distinctly different mixing enthalpies on either side of $x = 0.5$. To test this hypothesis, we plot $\Delta_{max}$ vs. the difference in metallic radii ($\delta R$) in Fig. 3(c). A moderate correlation is observed with a Pearson correlation factor of +0.715, wherein a value of +1 indicates a strong linear correlation, with a few notable exceptions. For example, Cr-containing pairs, such as Cr–Ti, Cr–Ta, and Cr–Nb, show relatively low asymmetry (~10 meV/atom) despite large differences in the radii (~20 pm). These systems are known to form stable Laves phases, and the size mismatch is a recognized driving force for intermetallic formation [37]. Therefore, even though the mixing enthalpy is not skewed, the size difference drives a different phase to become energetically favorable [38]. Excluding these exceptions, pairs containing Hf or Zr generally show both large differences in the radii and high asymmetry, reinforcing the link between these factors. In contrast, Ti-containing pairs (highlighted in bold in Fig. 3(b) and shown in gray in Fig. 3(c)) show smaller differences in radii and correspondingly lower asymmetry. The one exception is Cr-Ti for reasons discussed above. Taken together, these results suggest that a combination of lattice preference and atomic size mismatch influence the degree of asymmetry in the $\Delta H_{mix}(x)$ curves and change the phase boundary predictions. In contrast, 18 out of the 21 HCP pairs exhibit a much larger degree of asymmetry, as seen in Fig. 10 in Appendix B, since six of the nine elements prefer a BCC lattice. Notably, the Ti–Zr, Hf–Ti, and Hf–Zr binaries display negligible deviations between the regular and sub-regular fits, again demonstrating that elements sharing the same stable lattice, HCP in these cases, can be captured using the regular solution model.

The above analysis provides guidance for high-throughput computational workflows: sub-regular solution models should be employed for binary systems containing elements with differing stable lattices, whereas a regular solution model suffices for pairs where both elements favor the same lattice.

### 3.3. Benchmarking binary phase diagram predictions

Since our framework is based on binary mixing enthalpies, evaluating the accuracy of the predicted binary phase diagrams is a necessary first step before extending to multinary systems. Using the interaction parameters derived in the previous section, together with experimental unary data and intermetallic energies from the Materials Project database, we generated thermodynamic database (TDB) files as inputs for the *pycalphad* package. These files were then used to construct composition–temperature phase diagrams.

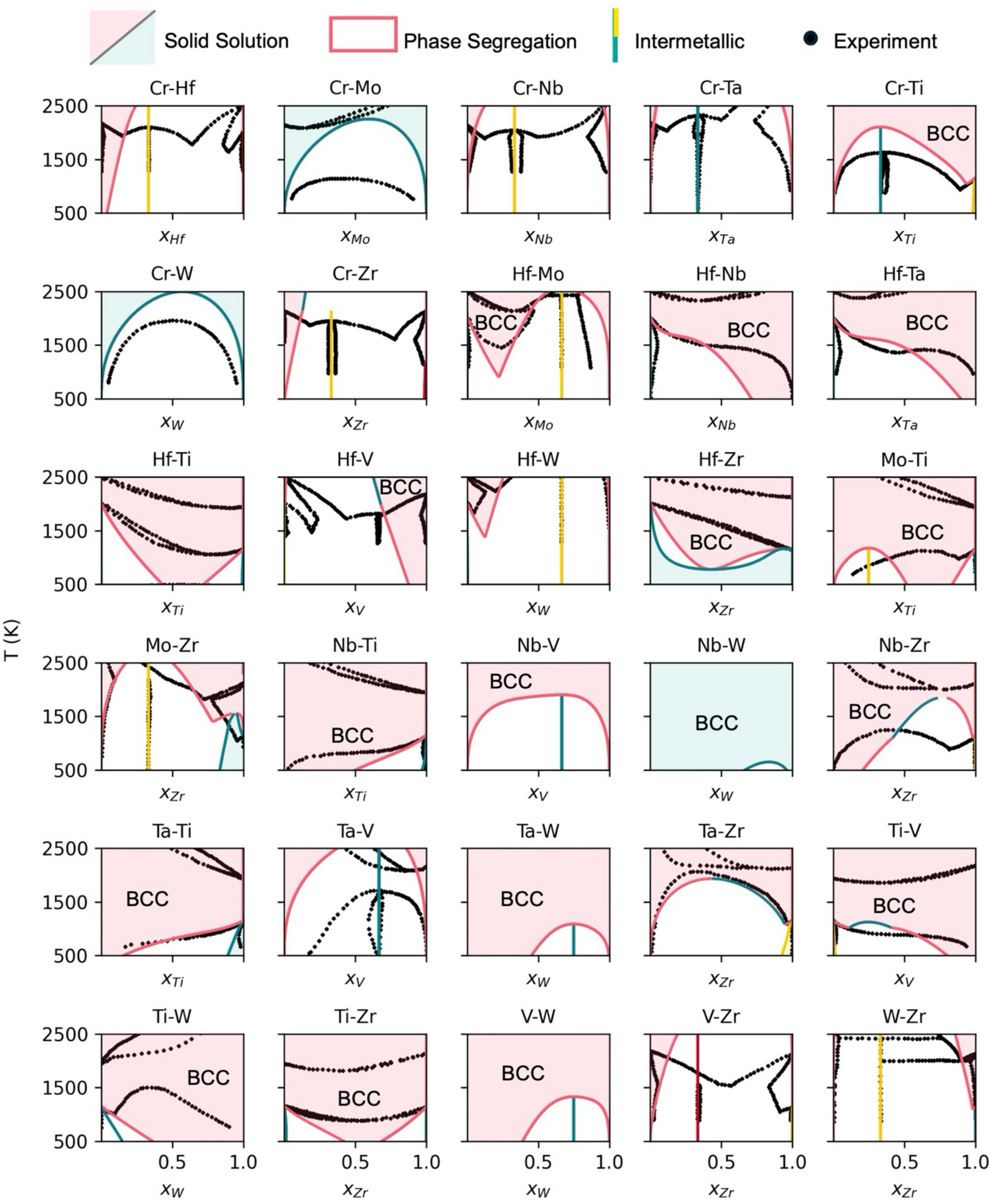

**Fig. 4. Comparison between experimental and predicted phase diagrams of 30 refractory binary alloy systems.** The red or green shaded regions depict the predicted BCC and HCP single phase solid solution regions, respectively, and vertical lines imply ordered intermetallics. The black dots are experimentally observed phase boundaries. Regions of phase segregation, either between two solid solutions or a solid solution and an intermetallic are shown in white.

Out of the 36 binary systems studied, predicted phase fields for 30 are shown in Fig. 4, along with experimental diagrams shown in black, where available. The majority of experimental diagrams are reproduced from Ref [1]. The Cr-W phase diagram is taken from Naidu et al. [39]. All the experimental plots were digitized and converted to a consistent scale of atomic fraction versus temperature (K) to ensure uniform comparison. The remaining 6 systems including Cr–V, Mo–Nb, Nb–Ta, Mo–Ta, Nb–W, and Mo–W were excluded from plotting since they are fully miscible across all temperatures and compositions, in agreement with the experimental results [1].

Despite its simplicity, our model can capture miscibility gaps and solid solution phase fields across the temperature and composition space in these binary phase diagrams. The predicted miscibility temperature, $T_{misc}$, at $x$ = 0.25, 0.50, and 0.75 are compared with experimental values in Fig. 5. The compositions for $x$ = 0.25, 0.75 are grouped into off-equimolar compositions, and $x$ = 0.5 equimolar composition is shown as a separate group. We first evaluate if the model can predict whether the alloy at a certain composition is miscible. Experimentally, a composition is classified as miscible if it is reported to form a single-phase solid solution below the experimental melting point, which is obtained from their binary phase diagram. The model prediction is then evaluated by asking whether the same composition and temperature are predicted to lie within a single-phase solid-solution field. From Fig. 5(a), it can be seen that for equimolar compositions (upper panel in red), the model's predictions match with the experimental observations for 32 out of the 36 systems, yielding an accuracy of 88%. Since 28 of the 36 equimolar compositions in the binary dataset are miscible, the dataset exhibits a substantial class imbalance. We therefore compare our results with a majority-class baseline that predicts all systems as miscible. This baseline gives an accuracy of 77.8% for the equimolar compositions, whereas the model gives a balanced accuracy of 83.9%, with true-positive and true-negative rates of 92.9% and 75.0%, respectively. Similarly, for the off-equimolar compositions, the model correctly classifies 67 out of 72 cases, giving an accuracy of 93.1%, compared with a majority-class baseline of 76.4%. The corresponding balanced accuracy is 91.4%, with true-positive and true-negative rates of 94.5% and 88.2%, respectively. For the systems showing miscibility below their melting point, we plot a histogram of the differences in experimental and predicted $T_{misc}$ in Fig. 5(b). The average error for both off-equimolar and equimolar compositions is ~500 K, with 50% of predictions lying within ±300 K, as seen from the median (orange line) of the distributions.

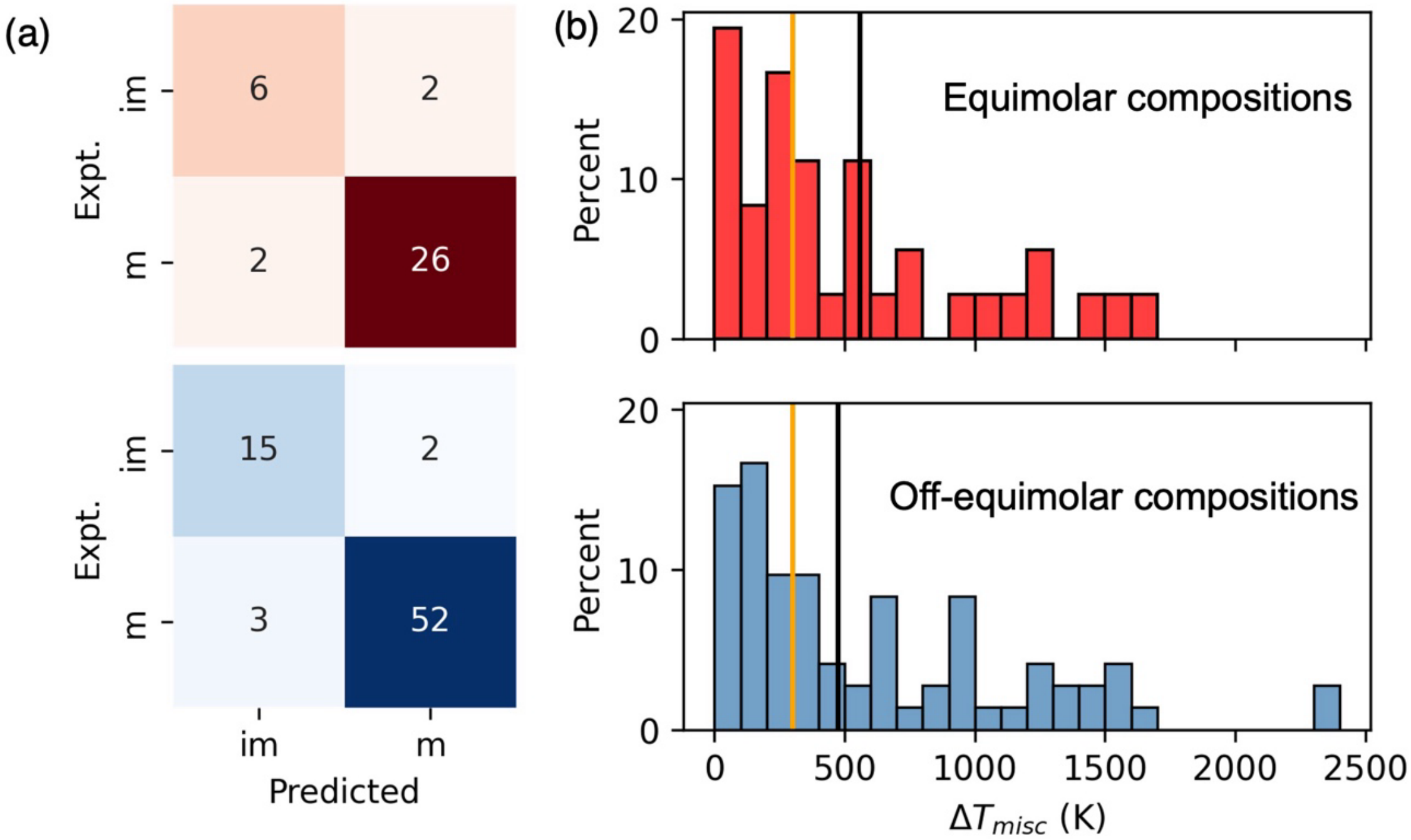

**Fig. 5. Experimental benchmarking of miscibility predictions for binary alloys across composition space.** (a) Confusion matrices comparing predicted and experimentally observed miscibility for equimolar ($x = 0.5$, top) and off-equimolar ($x = 0.75, 0.25$, bottom) compositions. Rows denote experimental outcomes, (miscible: *m*; immiscible: *im*), while columns denote model predictions. (b) Distributions of the predicted miscibility temperature difference, $\boldsymbol{\Delta T_{misc}}$, for equimolar (top) and off-equimolar (bottom) compositions. The black vertical line indicates the average error and orange vertical line denotes the median.

Additional observations can be made from the qualitative comparison between the predicted and experimental phase fields shown in Fig. 4. The model overestimates $T_{misc}$ for a few BCC+BCC systems, most evidently in Cr-Mo and Cr-W. The $T_{misc}$ of the BCC phase for BCC+HCP systems, such as Cr-Hf, Nb-Hf, Nb-Ti, Ta-Hf, is underestimated at higher compositions of Cr, Nb, Nb, and Ta, respectively. The Ti-W system shows the largest discrepancy between the predicted and experimental phase diagrams. Ångqvist et al. [36] used a cluster expansion model fitted to DFT energies to predict the phase diagram of the Ti-W system. They found a positive mixing enthalpy for the HCP lattice, which could not be reproduced in this study. The reason for a miscibility gap comes from the previously discussed asymmetry in the Ti-W BCC mixing enthalpy, which the sub-regular model does not capture, as well as the effects of anharmonic phonons. Thus, our model overestimates the stability of the BCC phase across all compositions in the Ti-W alloy system. Similar over-stabilization of the BCC phase is observed in Hf–Ti, Ti–Zr, and Hf–Zr, where the HCP-to-BCC transition is known to be driven by anharmonic vibrational entropy [29]. Thus, anharmonic phonon contributions are critical to accurately describe these systems. Ångqvist et al. [36] approximated the energy contributions of the anharmonic phonons for the binary alloy by extrapolating the contributions calculated from unary Ti. Even for unary calculations, ab-initio molecular dynamics calculations have to be performed at elevated temperatures to fit the energy contributions from phonons [40, 41]. While these effects can be incorporated into the model, doing so would require significant computational effort.

Apart from the prediction of phase-segregated regions in the temperature-composition phase space, the prediction of stable intermetallics is also an important aspect of these phase diagrams. The Materials Project database [27] contains an extensive list of binary intermetallics. All but two of the intermetallics seen experimentally in the thirty binary systems are predicted to be stable by DFT. Notably, the refractory systems considered here, such as Cr-Hf, Cr-Nb, Cr-Ta, Cr-Ti, Cr-Zr, Hf-Mo, Hf-W, Mo-Zr, Ta-V, V-Zr and W-Zr, host cubic Laves phases, which are both predicted to be stable and experimentally confirmed, as can be seen from Fig. 4. The high-temperature hexagonal Laves phases are not included here, as temperature effects on intermetallic free energies are not explicitly modeled. However, these phases can be incorporated into our model in a relatively straightforward manner, though doing so would require significant computational effort. As an example, the room temperature free energy of the C15 Laves phase of $ZrV_2$ was calculated from its phonon density of states. The cubic Laves phase of $ZrV_2$ becomes stable at ~110 K [28], and hence is not predicted to be stable by DFT calculations at 0 K. The phase stability as a function of temperature for $ZrV_2$, as well as the reproduction of its heat capacity and transition temperatures are provided in Appendix C. $HfV_2$, analogous to $ZrV_2$, is also unstable at 0 K, but can be incorporated similarly. However, intermetallics in other systems can be more complex to model, such as in the Ta-V binary phase space. $TaV_2$ is first stable as a C14 Laves phase at room temperature and then stable again at high temperatures, competing with the cubic Laves phases [42]. Thus, our model overestimates the stability of the C15 $TaV_2$ phase leading to a phase segregated region that stretches beyond the experimental liquidus temperatures. Experimentally, up to ~20% of V by atomic fraction is soluble in Ta at room temperature, but we predict phase segregation at any amount of V. Thus, the stability of intermetallics not only determines the miscible temperature of solid solutions, but also the *compositional span* of the miscible region. These examples highlight the need for explicit treatment of vibrational free energies of intermetallics in binary systems to better reproduce experimental observations—a feature we identify as a key direction for future development.

In some systems, our model predicts ordered intermetallics that have not been experimentally reported. A previous computational report predicted a $T_{misc}$ at 1250 K for the equimolar alloy in the Nb-V system [11], but no stable intermetallics were reported. However, $NbV_2$ Laves phase was predicted to be stable by Li et al. [43] and a ternary laves phase, $(Nb, Zr)V_2$, was experimentally observed. This Laves phase, not included in the Materials Project database, is added to our pool of intermetallics, as shown in Fig. 4. Similarly, in the Mo-Ti system, no stable intermetallics are experimentally observed; however, previous computational reports predict that Mo-Ti has a high propensity for ordering and hosts multiple stable intermetallic phases [44], with one of them being the $TiMo_3$ phase predicted to be stable by our model too. Similarly, the intermetallics predicted in the Ta-W and V-W systems have not been experimentally observed. Previous studies on V-W alloys have primarily focused on the properties of the high-temperature phases and their melts, with limited attention to low-temperature phase stability. For the Ta-W system, Turchi et al. [45] predicted a stable equimolar B2 phase, whereas the Materials Project reports $TaW_3$ to be more stable. Our model captures that Ta-W will stabilize into an intermetallic at lower temperatures but the competition between the two phases, $TaW_3$ and $Ta_1W_1$, needs to be examined further. Taken together, these results show that our method reproduces the majority of binary phase fields, capturing both miscibility gaps across temperature and composition, as well as intermetallic formation; two factors that determine the compositional span of single solution regions. At the same time, they highlight areas—particularly temperature-stabilized intermetallics and systems requiring vibrational free energy corrections—where future improvements will enhance accuracy and predictive power.

### 3.4. Comparison of predicted phase diagrams of ternary refractory alloys with experiments

We next benchmark predicted ternary phase diagrams against available experimentally reported isotherms. Eight isotherms from four ternary systems are reported in Ref. [46]. Comparisons of the experimental and predicted phase diagrams at different temperatures are shown in Fig. 6, and in Fig. 12 in Appendix D. For the Cr–Ti–V system, the predicted phase boundaries align well with experimental observations, as shown in Fig. 6(a). However, because our model does not include temperature-stabilized intermetallics, it fails to predict the hexagonal $TiCr_2$ Laves phase. In the Cr–Nb–W system, our model cannot reproduce the mixing of Nb and W within the $(Nb,Cr)W_2$ Laves phase, as can be seen in Fig. 11(a), because we don't consider the enthalpy of mixing in sublattices of intermetallics. Determining the free energy of complex intermetallics requires further computational effort and is a key future direction, as they change the phase boundaries quite significantly as seen in Ta-V and V-Zr binaries, and others discussed below. Moreover, the overestimation of the Cr–W miscibility gap, as previously discussed, leads to an exaggerated region of phase segregation for Nb–W–rich compositions. We plot the predicted isotherm at 2400 K, while the experimental observation is at ~1800 K to highlight the underestimated phase stability of the solid solution. Thus, while we predict the right trend in solid solution regions for some systems, the errors from the binaries lead to a change in the predicted miscibility temperatures for their ternary alloys. A similar trend is observed in the Cr–Mo–W system, where the Cr–Mo miscibility gap is overestimated; however, at higher temperatures, our model correctly predicts an expansion of the miscible phase field in the Cr–Mo–W ternary, as shown in Fig. 6(c), where we compare the experimental isotherm at ~1600 K with the one predicted at 2300 K. For the Mo–Nb–Ti system, Senkov et al. [47] have reported that the equimolar alloy is stable after annealing at 1473 K, whereas the experimental phase diagram from Ref. [46] indicates phase segregation between (Mo,Nb)-rich and Ti-rich phases at the same temperature. Our predictions show nearly complete miscibility across the phase space above 800 K (see Fig. 11(b)) aligning well with recent experimental observations [47]. Zhang et al. [12], also predicted the equimolar MoNbTi alloy to be stable, consistent with our results.

Our model predictions are in good agreement with reported phase behavior in equimolar ternary alloys. Zyka et al. [48] and Hu et al. [49] reported that the equimolar NbTaTi and NbTiZr alloys form single-phase solid solutions. Both results are consistent with our predictions of stable single-phase solid solutions above 1000 K. Li et al. [43] reported phase segregation in the equimolar NbVZr alloy, with a BCC + Laves microstructure. Our predicted phase diagram for NbVZr also shows BCC + Laves phase segregation at 1000 K, in good agreement with experiments.

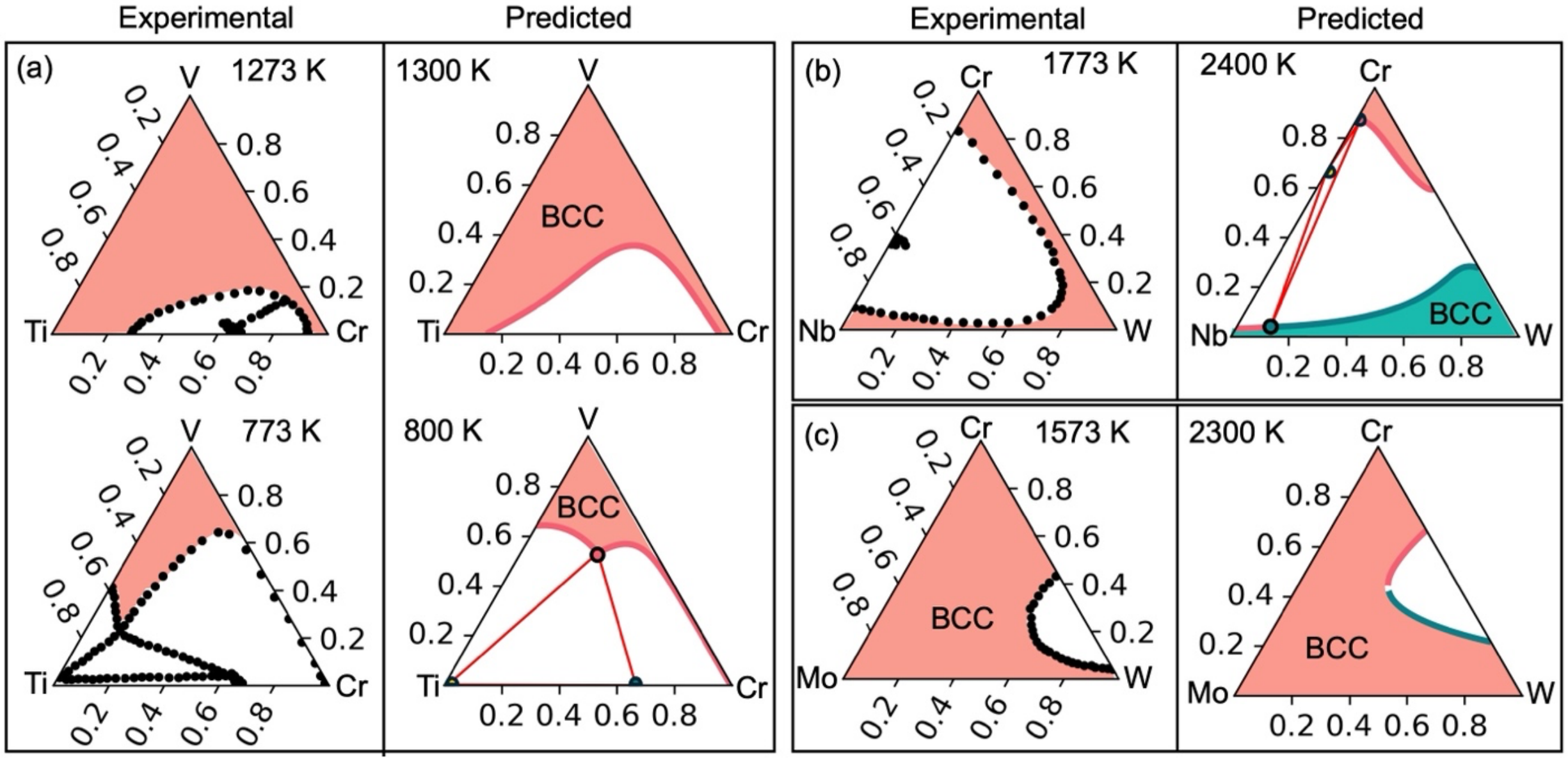

**Fig. 6. Comparison between predicted and experimental ternary isotherms.** (a) Cr-Ti-V at 773 K and 1273 K, (b) Cr-Nb-W at 1773 K, (c) Cr-Mo-W at 1573 K. Note that the predicted isotherms are at elevated temperatures for (b) and (c), highlighting the underestimation of the stability of the solid solution. Shaded regions in the isotherms refer to single phase solid solution regions and white regions exhibit phase segregation. Colored circles imply invariant points.

Apart from the experimental reports, we also compare our predictions with those from the commercially available PANDAT database [16] for ternary isotherms at 1500 K. Six ternary systems are chosen randomly and presented in Fig. 7, and three more in Fig. 13 in Appendix D. The rest of the predictions are available in the publicly available repository [26] and can be explored interactively at the web-based interface accompanying this article. Many of the ternary systems that do not compare well with PANDAT classifications are due to the two primary reasons discussed earlier. Firstly, the lack of data for ternary intermetallics, particularly, data on the sublattice mixing enthalpies for the Laves phases results in the prediction of three-phase segregation rather than two-phase segregation, as predicted by PANDAT, and shown in Fig. 7(b)-(c) and Fig. 12(b). These partially disordered ternary intermetallics, which can be stabilized by the configurational entropy, change the phase boundaries. A study on the stability of these multinary intermetallics will be the focus of a future work. Secondly, errors originating in the binary systems propagate into the ternary predictions. Particularly, ternary systems containing Cr-Mo have an overestimated $T_{misc}$, and systems containing Ti-W underestimate the $T_{misc}$. Thus, predictions from higher order systems containing these two binary pairs should be considered with care. We expect isotherms calculated at higher temperatures to more closely reproduce lower-temperature experimental observations for Cr-Mo, whereas the opposite trend is anticipated for Ti-W. Apart from these issues, the model is able to predict the phases in equilibrium and regions for finding stable solid solutions quite accurately for the majority of the ternary alloy systems, as shown in Fig. 7 and Fig. 13.

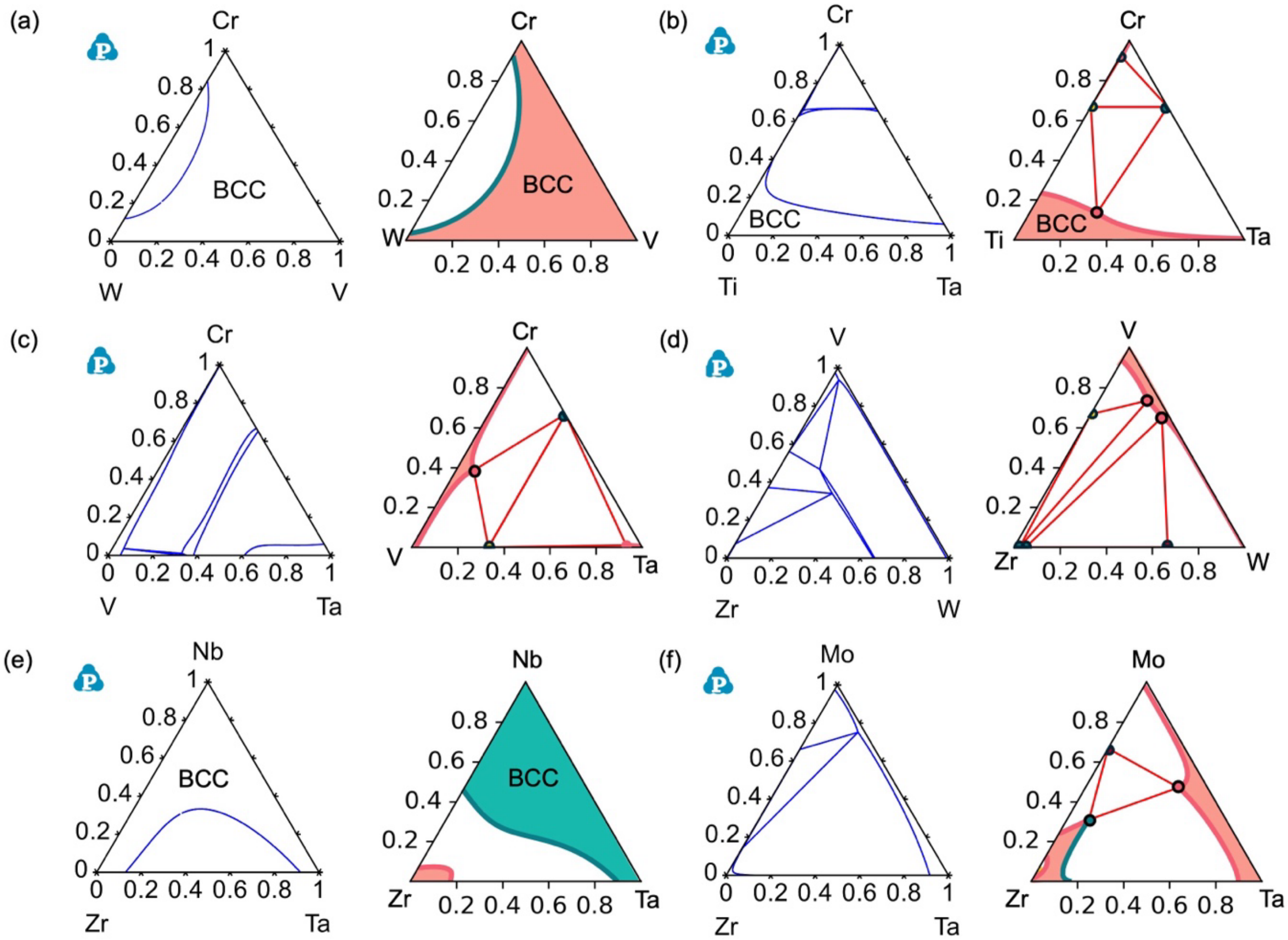

**Fig. 7 Comparison between PANDAT-predicted and our model-predicted ternary isotherms at 1500 K.** (a) Cr-V-W, (b) Cr-Ta-Ti, (c) Cr-Ta-V, (d) V-W-Zr, (e) Nb-Ta-Zr, (f) Mo-Ta-Zr. Shaded regions in the model-predicted isotherms show fields with single-phase solid solutions and white regions indicated phase segregation. Colored circles mark invariant points.

### 3.5. Comparison with existing first-principles frameworks

The model discussed in this work has both advantages and limitations compared to existing first-principles-based approaches for alloy phase stability prediction. The central distinction is the balance between atomistic detail and computational tractability.

The approach we have outlined above is conceptually similar to the *sqs2tdb* framework developed by Van de Walle et al. [31], where the authors proposed generating SQS structures at multiple compositions and for systems with increasing component count, followed by obtaining mixing enthalpy and vibrational free energy from DFT calculations along with contributions from short-range order approximated using the cluster variable method. Temperature-dependent properties could then be obtained through phonon calculations, while liquid energies were estimated via *ab-initio* molecular dynamics (AIMD) simulations. These contributions were fit using Redlich–Kister polynomials [50], and the CALPHAD formalism was used to construct the phase diagrams. While rigorous, this framework becomes computationally intractable for high-throughput applications of MPEAs due to the number of *ab-initio* calculations that need to be performed.

A similar comparison can be made with the multi-cell Monte Carlo (MCMC) approach developed by Niu et al. [51]. MCMC uses multiple simulation cells to sample phase coexistence, ordering, and chemical partitioning in complex alloys. The method provides a more explicit atomistic description of phase separation than the present model with information on short and medium-range ordering, but its reliance on configurational sampling and repeated energy evaluations again makes it better suited for detailed studies of selected systems rather than screening across large alloy libraries. Cluster-expansion-based Monte Carlo approaches also provide rigorous descriptions of high entropy alloys, but are not logistically feasible for higher-order alloys because they require thousands of atomistic calculations [52].

Yang et al. [53] developed an *ab-initio* framework in which unique supercell configurations of nearly equivalent stoichiometry are enumerated and combined using statistical averaging to estimate finite-temperature properties of partially occupied or disordered materials. While it provides a systematic treatment of configurational disorder and off-stoichiometry, it requires explicit enumeration and first-principles evaluation of many configurations for each system, as modelled within the AFLOW framework.

By contrast, our method does not attempt to explicitly calculate every composition, configuration, or temperature-dependent free-energy contribution. Instead, it uses composition-dependent sub-regular solution models, supplemented by first-principles intermetallic formation energies, to construct thermodynamic descriptions of higher-order alloy systems. The computational advantage comes from separating the upfront binary database construction from the marginal cost of generating higher-order phase diagrams. For the nine-element refractory space considered here, the solid-solution database requires DFT calculations only for the constituent binary interactions: corresponding to ~300 disordered solid-solution calculations and 36 ordered-compound calculations. Once these binary interaction parameters and ordered-compound formation energies are available, no additional ternary, quaternary, or quinary disordered SQS calculations are required to generate higher-order phase diagrams. For example, a quinary alloy contains ten binary pairs; constructing a new quinary database from scratch would therefore require only the binary calculations associated with these ten pairs, rather than explicit first-principles calculations of the full quinary configuration space. If the constituent binaries are already present in the database, the marginal first-principles cost of generating a new quinary phase diagram is zero. In contrast, a typical Monte Carlo run for estimating the energy of a single quinary alloy composition would require hundreds of atomistic calculations. With the method presented in this work, each binary phase diagram or ternary isotherm can be generated within 10s. Quaternary and quinary phase diagrams are generated by predicting the phase information for 200-500 compositions and require 20-30s for computation (as checked on the web interface)—effectively requiring ~1s per composition per temperature. The phase diagrams can be generated in the publicly available web interface here: https://raptor.engr.wustl.edu . Descriptions of the modules available in the web-interface and their functionality are provided in Appendix E.

It should be noted that the other methods can predict properties not accessible by our model in its current form, like microstructure, grain size, strain effects and short-range order. However, our model is flexible to accommodate the effects of vibrational entropy, magnetic entropy and short-range order for systems that require the additional accuracy. Our model is not intended to replace detailed CALPHAD assessments, cluster-expansion studies, or Monte Carlo simulations of individual alloy systems. Rather, it provides a computationally efficient screening framework—providing sufficiently reliable predictions of solvus phase boundaries in MPEAs, as discussed in Part 2, while remaining computationally tractable for high-throughput exploration. In this sense, the approach occupies a balance between the efficiency of empirical alloy-design rules [54, 55] and accuracy of fully atomistic thermodynamic calculations outlined above in this section.

## 4. Conclusions

We have refined and benchmarked a computationally efficient framework for predicting temperature–composition phase diagrams of refractory alloys using DFT-derived binary mixing enthalpies combined with sub-regular solution models. By representing mixing enthalpies as continuous functions of composition and explicitly incorporating temperature-dependent unary reference states, the framework resolves key shortcomings of convex-hull-based approaches [12-14], including an underestimation of the miscible temperatures of solid solutions. The integration of the DFT-calculated enthalpy with the open-source *pycalphad* package enables high-throughput generation of phase diagrams and provides a flexible interface for comparison with, and integration into, existing commercial thermodynamic databases.

Benchmarking against 36 binary and 15 ternary systems shows that the model reliably captures miscibility gaps, solid-solution stability, and intermetallic formation, with predicted miscibility temperatures lying within ±300 K of experimental values. Our analysis further elucidates that regular solution models are adequate for binaries whose end members share the same stable lattice, whereas sub-regular models are necessary for systems involving different ground-state lattices. These insights provide practical guidance for constructing high-throughput thermodynamic models for higher-order alloys with minimal additional computational cost, which is the focus of Part II.

While the present framework does not explicitly include vibrational free energies of solid solutions or temperature-stabilized intermetallics, the results show that a binary-based solution-model approach can reproduce key qualitative and quantitative features of experimental phase diagrams for the majority of binary systems considered here. We also identify that temperature-dependent phase stability of intermetallics, particularly in Hf-V, Zr-V, and Ta-V systems, as well as sublattice mixing in ternary intermetallics, are important features to incorporate in future work. We further demonstrate, using the Zr-V system as an example, that the effects of vibrational energy can be incorporated in a straightforward manner by either phonon calculations or by using database-derived phonon data when available [56].

In Part II of this work, we extend this methodology to higher-order systems, benchmark the predictions against a broader experimental dataset, analyze phase-stability trends across multinary temperature-composition spaces and formulate design rules to tune the microstructure of complex alloys. Together, these advances establish a scalable framework for phase-diagram prediction, supporting the accelerated design of refractory MPEAs.

**Data Availability**

A python-based package to analyze the MPEA phase diagrams predicted in this work has been provided in https://github.com/Materials-Modelling-Microscopy/RAPTor along with nearly 400 TDB files. Code to plot phase diagrams, compute miscibility temperatures and parse TDB files are also provided. The DFT calculations are provided in a publicly available Zenodo repository [26]. A web interface is also made available to obtain predicted phase diagrams and related analysis for the alloys studied in this work at https://raptor.engr.wustl.edu .

**Declaration of Competing Interest**

There are no competing interests to declare.

**Acknowledgements**

This work was partially supported by the Schmidt Family Foundation (P.O., N.C., J.C., A.C., R.M.) and the National Science Foundation through award # DMR-2145797 (P.O., R.M.). This work used computational resources through allocation DMR160007 from the Advanced Cyberinfrastructure Coordination Ecosystem: Services & Support (ACCESS) program, which is supported by NSF awards #2138259, #2138286, #2138307, #2137603, and #2138296.

## Appendix A. Use of denser grids for free-energy minimization

As discussed in Section 3.1., the use of denser grids for the free-energy minimization leads to a noticeable difference in the $T_{misc}$. The reason for the underestimation of the $T_{misc}$ is evident in lower-order systems. As an illustrative example, a schematic binary convex hull with only one equimolar solution is depicted in Fig. 8(a) at a temperature of 2000 K. A denser grid, consisting of several non-equimolar compositions, is used to build the convex hull in Fig. 8(b). It can be clearly seen that the equimolar composition is on the hull in the first approach but is unstable in the second approach. Thus, the use of denser grids will more accurately estimate the $T_{misc}$ for multinary alloys. We calculate the convex hull using the approach implemented in the pymatgen package [57], with a denser grid containing of 15 points in any composition direction, and predict the $T_{misc}$ for binary, ternary, quaternary and quinary equimolar compositions. Then we compare the predictions with that of a coarser grid consisting of only equimolar compositions. The histogram of the differences between the two sets of $T_{misc}$ are plotted in Fig. 8(c). The average difference in temperature across all orders, is positive, indicating the above-mentioned underestimation of the $T_{misc}$ by the coarse-grid approach. The mean difference increases slightly as more components are added, from 125 K to 250 K, however the tail of the distribution extends up to 750 K for the quinary alloys, thus making it more important to utilize a denser grid for multinary systems.

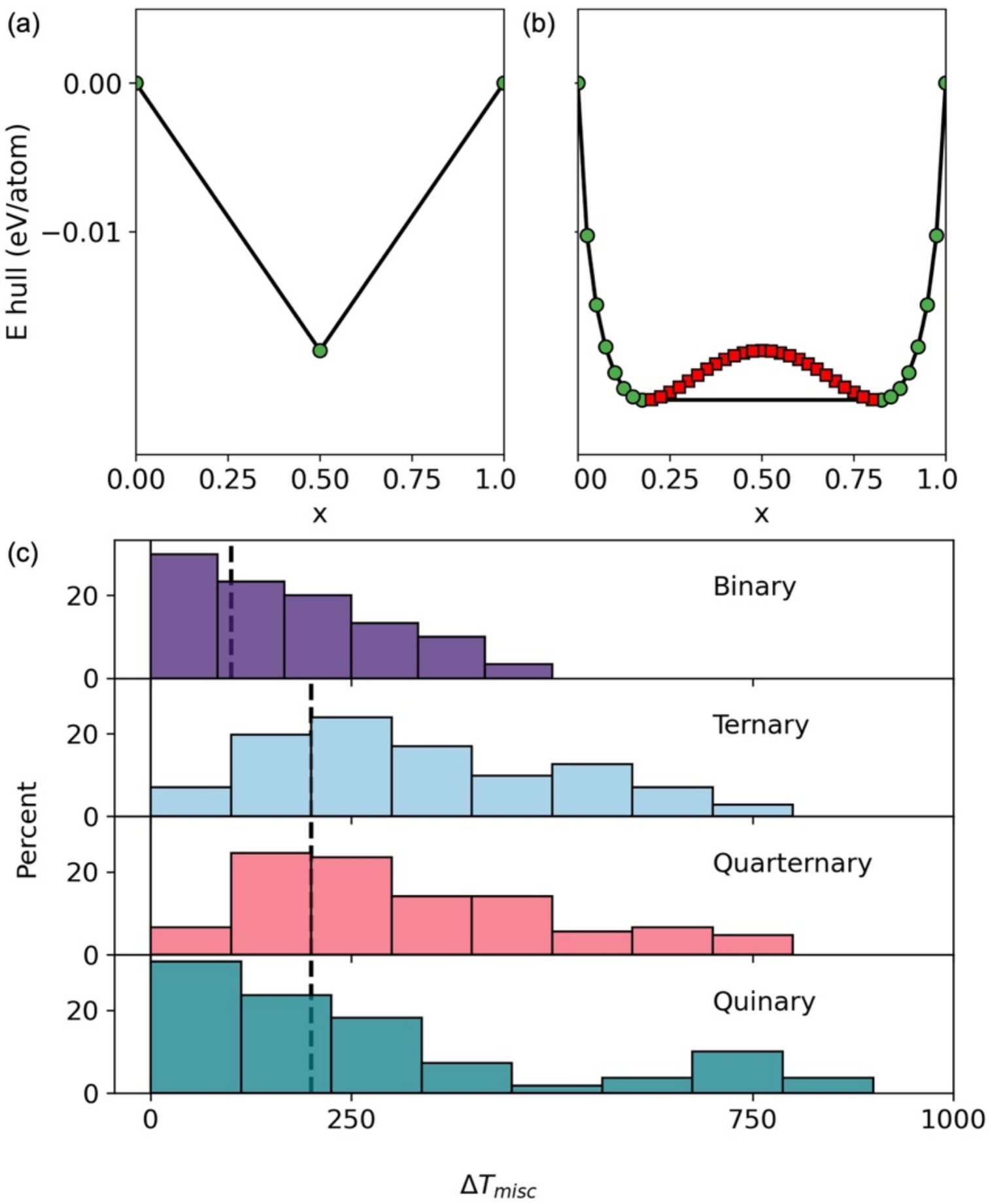

**Fig. 8. Effect of grid size on phase stability.** Two convex hulls were constructed from the Cr-W system at 2000 K, with a) Only an equimolar point used for constructing the binary convex hull, along with the elements energy set to 0, and b) using a denser grid approximating the Gibbs free energy derived from the solution models. c) The change in the $T_{misc}$ when a denser grid is used compared to the coarser grid.

**Appendix B. Solution-model fits to the mixing enthalpies of binary pairs**

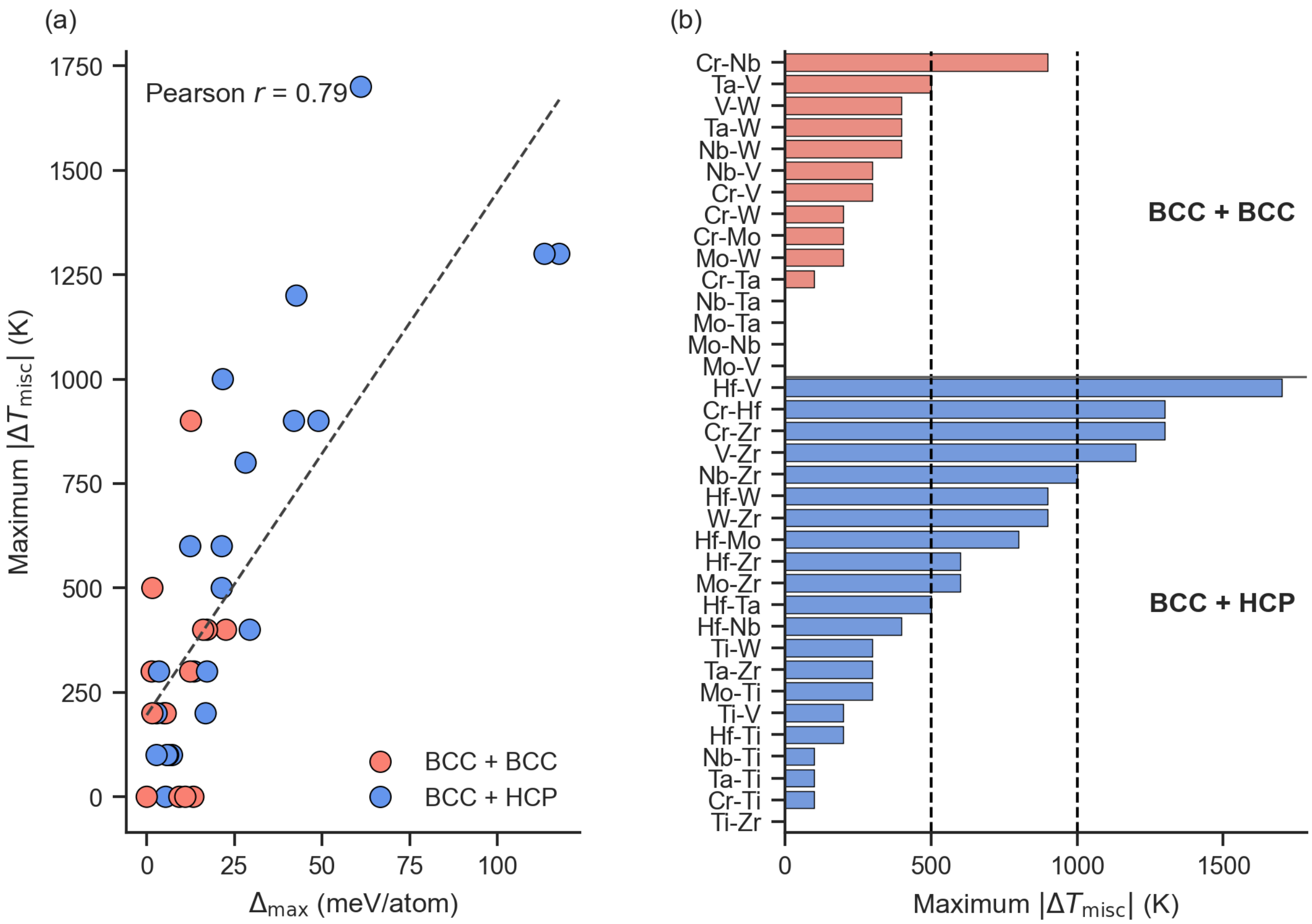

**Fig. 9. Phase-boundary sensitivity to regular and sub-regular solution descriptions.** (a) Maximum change in the predicted miscibility temperature, $\Delta T_{misc}$, obtained by comparing regular and sub-regular solution models at x=[0.25, 0.50, 0.75], plotted against the maximum energetic deviation, $\Delta_{max}$, between the two mixing-enthalpy descriptions. Points are colored according to whether both end members are BCC-stabilizing elements or whether the binary contains an HCP-forming element. The dashed line is a linear fit to all systems. (b) System-resolved values of maximum $|\Delta T_{misc}|$ grouped by lattice class.

DFT calculations were performed to obtain mixing enthalpies for 21 HCP binary alloy systems, specifically those containing Ti, Zr, or Hf, which were subsequently fitted using regular and sub-regular solution models. The fits are shown in Fig. 10.

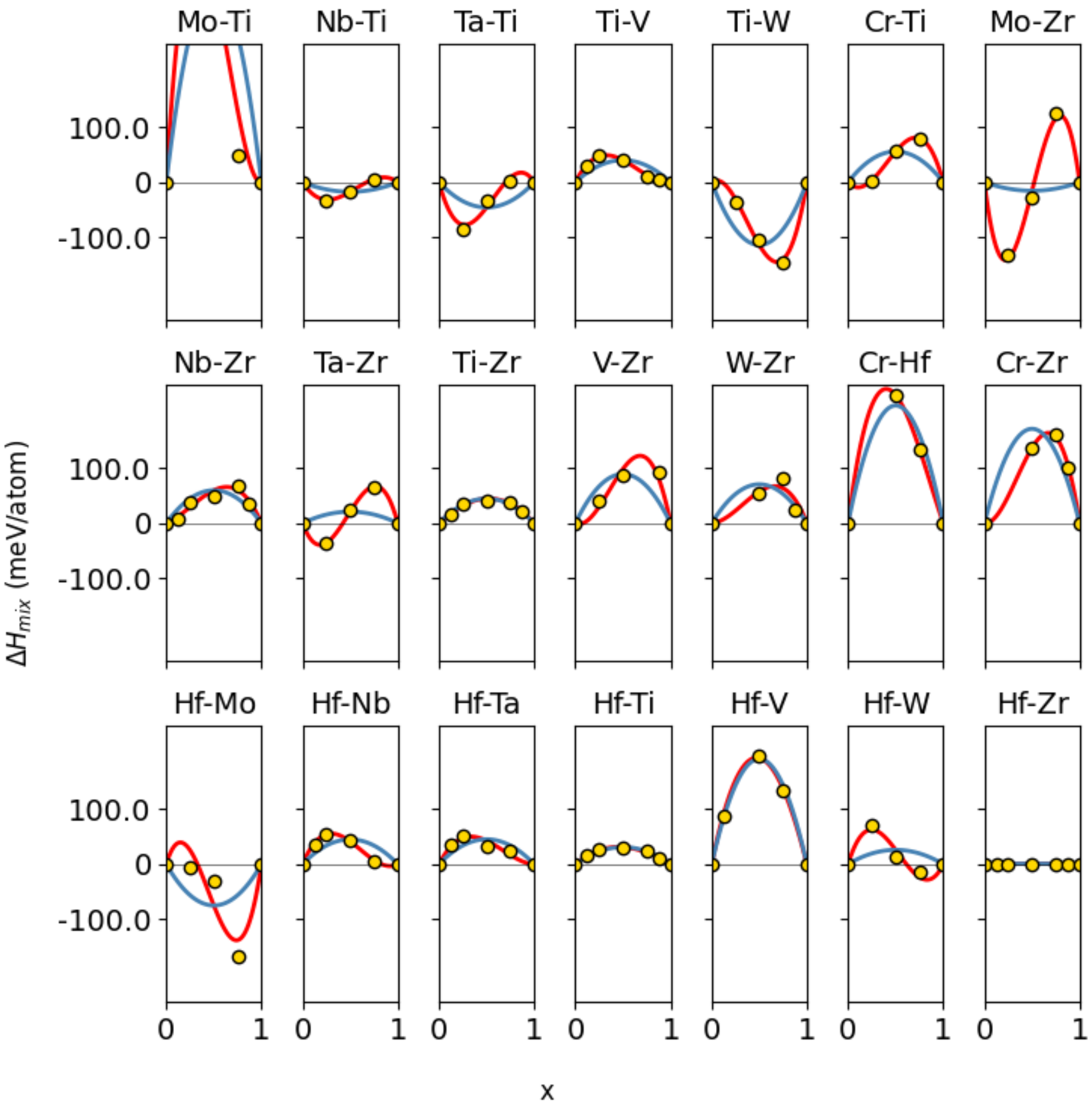

**Fig. 10. Composition-dependent mixing enthalpy $\Delta H_{\text{mix}}$ for the HCP lattices of the 21 binary systems that contain either Ti, Hf or Zr.** Yellow markers indicate DFT calculations. Red and blue lines indicate sub-regular and regular solution model fits, respectively.

**Appendix C. Vibrational free energy for $ZrV_2$ Laves phase**

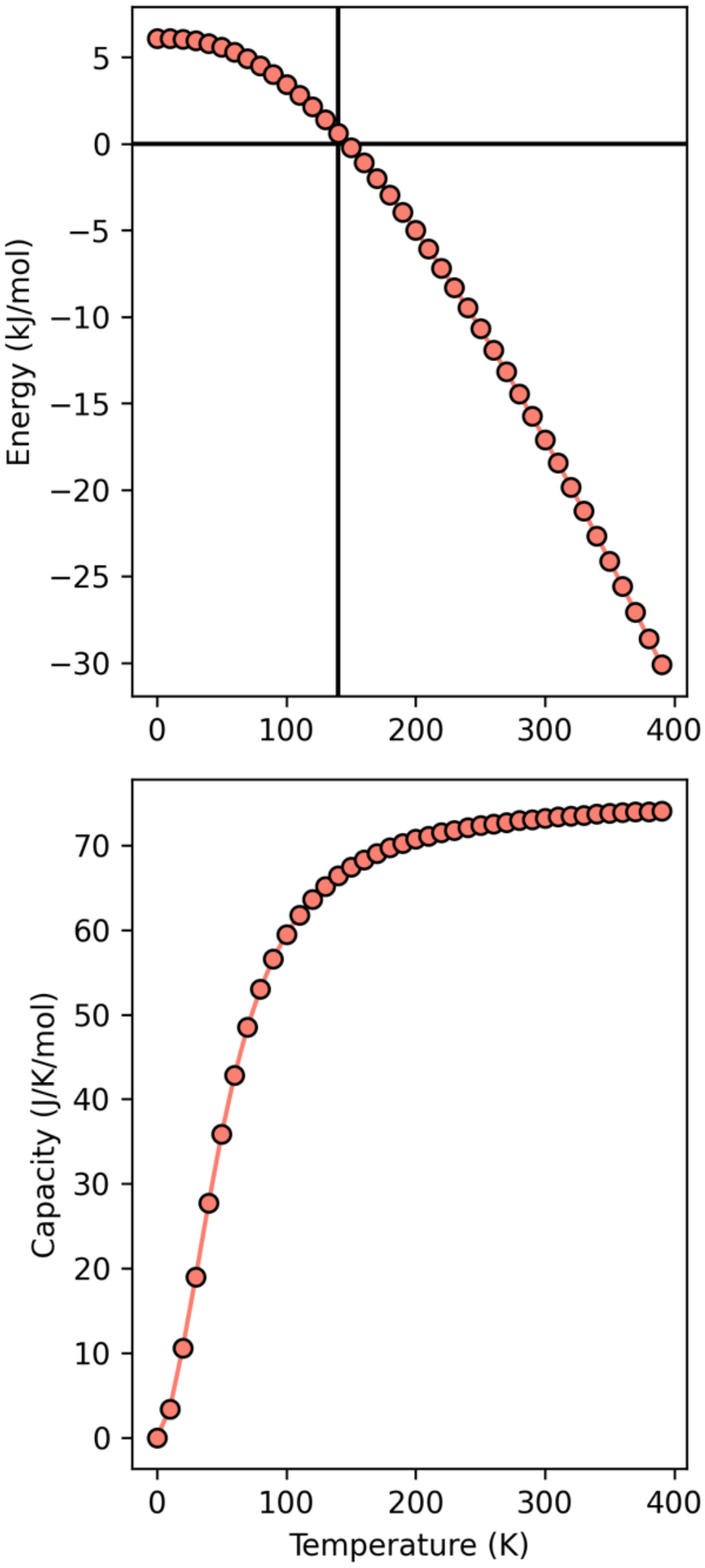

**Fig 11. Temperature-dependent properties for $ZrV_2$ Laves (C15) phase.** The free-energy is plotted against temperature in the top panel, and heat capacity in the bottom panel. The phase becomes stable at ~140 K, matching well with the experimental transition temperature of ~110 K [28].

**Appendix D. Comparisons between predicted and experimental/PANDAT ternary isotherms**

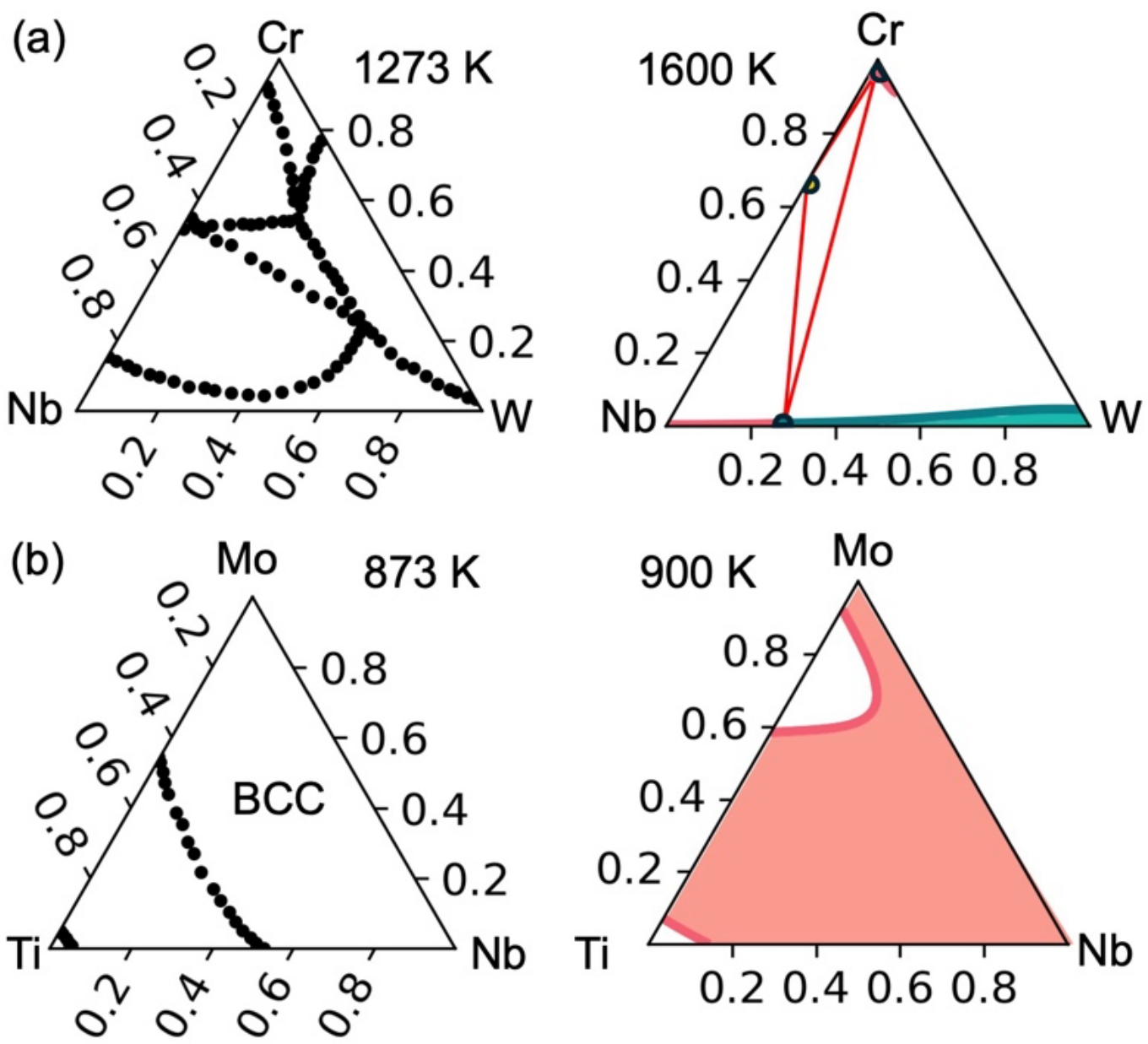

**Fig. 12. Ternary phase diagram comparison between experimental and model predictions at specific temperatures.** (a) Cr-Nb-W observed experimentally at 1273 K and the predicted isotherm at 1600 K. (b) Mo-Nb-Ti predicted experimentally at 873 K and the predicted isotherm at a similar temperature. Shaded regions in the predicted isotherms refer to single phase solid solution regions and white regions are exhibiting phase segregation. Colored circles denote invariant points.

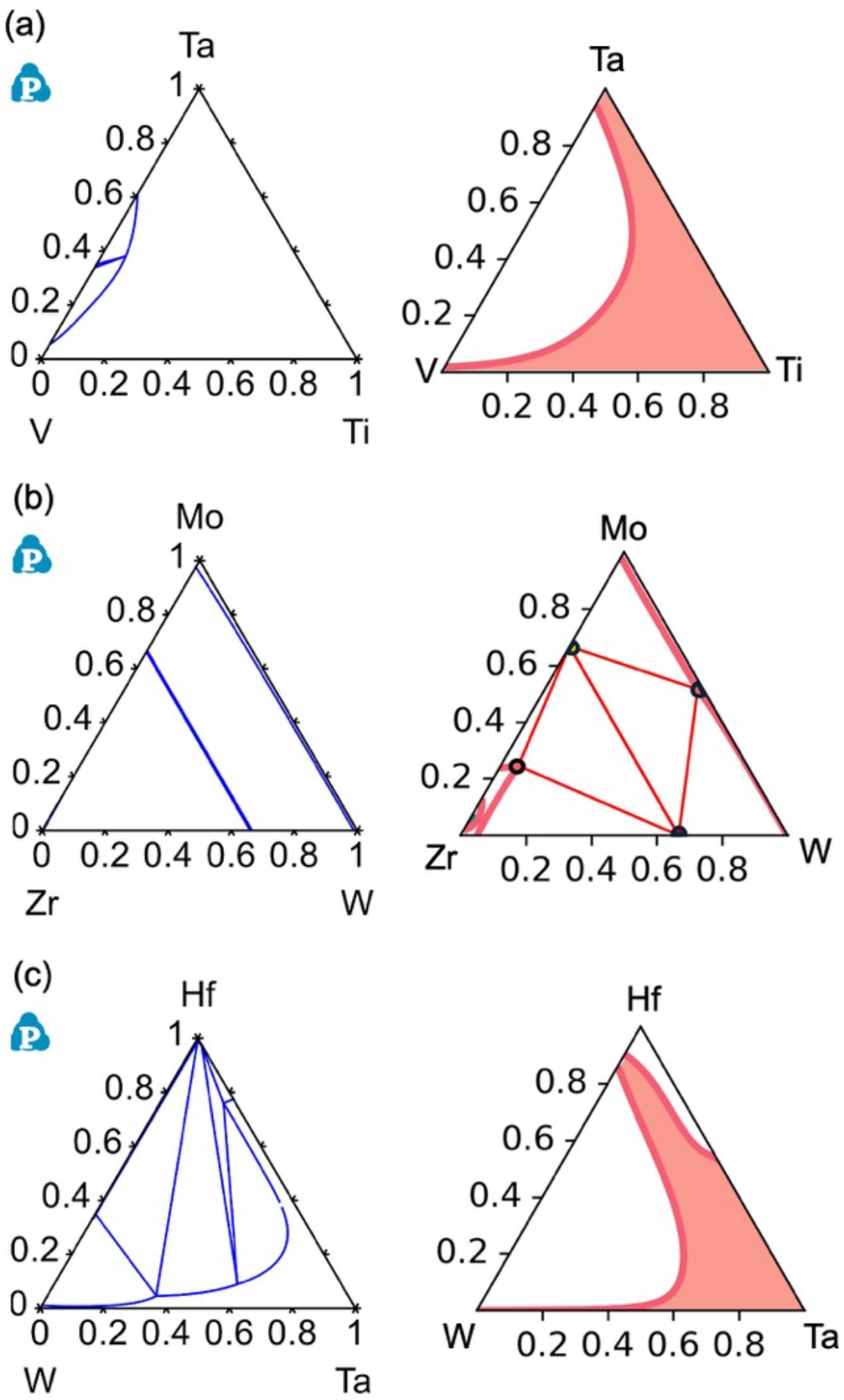

**Fig. 13. Ternary phase diagram comparison between experimental and model predictions at 1500 K.** (a) Ta-Ti-V, (b) Mo-W-Zr and (c) Hf-W-Ta. Shaded regions in the predicted isotherms refer to single phase solid solution regions and white regions are exhibiting phase segregation. Colored circles denote invariant points.

## Appendix E

To make the phase prediction framework accessible to users, we have developed a web-based interface for rapid alloy phase-diagram prediction and visualization, which is available at: https://raptor.engr.wustl.edu . The interface allows users to select an alloy system, temperature range, composition, and visualization mode, and then directly generate binary, ternary, and higher-order phase-stability maps. For binary systems, the tool produces temperature–composition phase diagrams, while ternary systems can be visualized as isothermal ternary sections. For fixed multicomponent compositions, the interface calculates phase fractions as a function of temperature, enabling direct assessment of predicted phase evolution during thermal processing, shown as snippets of the web interface in Fig. 14. For quaternary and quinary systems, the interface integrates the SymPlex visualization framework [58] to display composition-dependent properties such as solid-solution phase fraction and the number of predicted phases at any selected temperature. Users can also download the generated figures and underlying numerical data for further analysis. Overall, the web-interface serves as an interactive front end to the thermodynamic prediction framework, enabling rapid assessment of alloy phase stability across composition and temperature space.

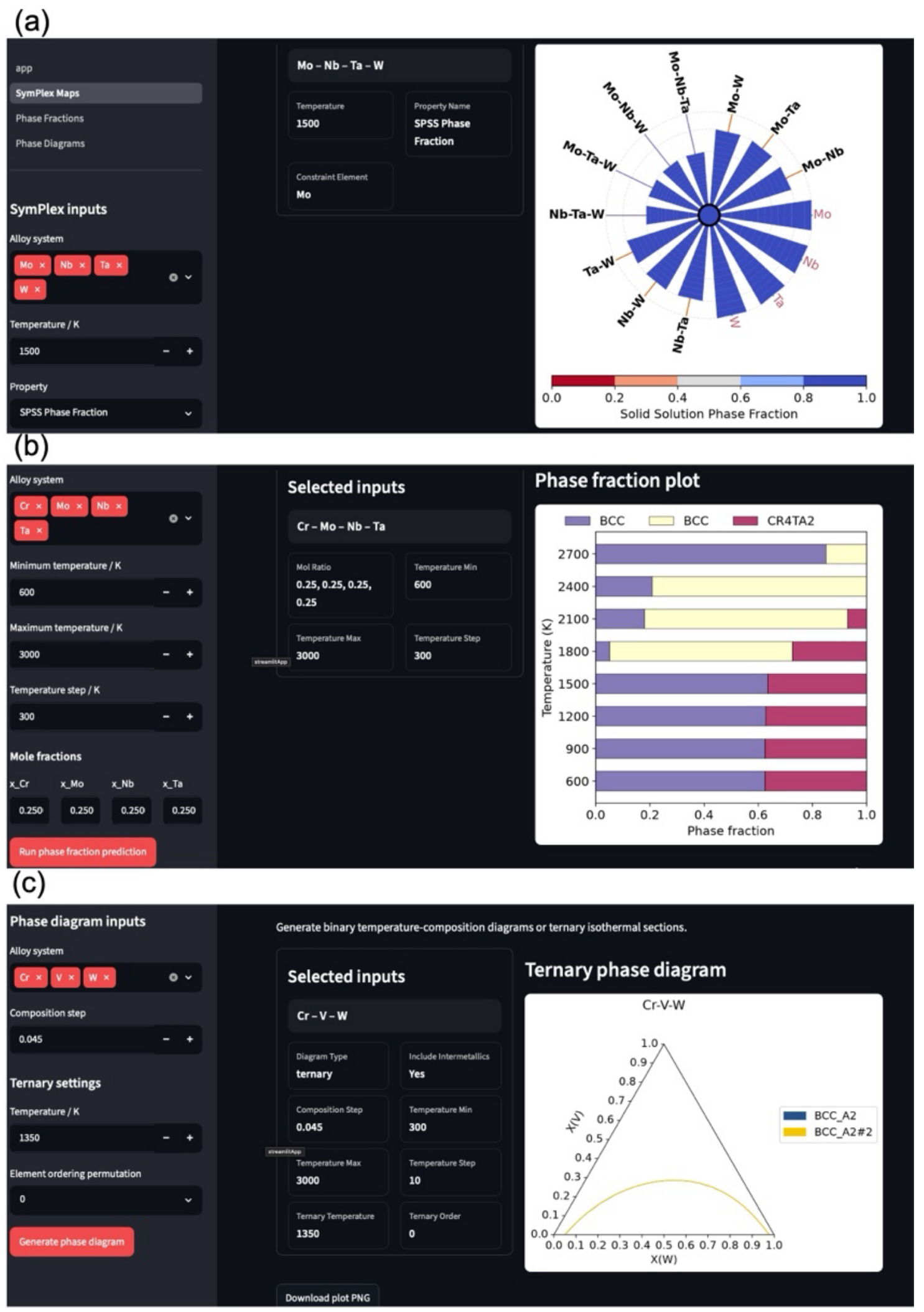

**Fig. 14. Web interface for rapid phase-field prediction and visualization. Representative screenshots of the deployed web interface showing the three main prediction modes.** (a) SymPlex visualization module for quaternary and quinary alloy systems, where users select the alloy system, temperature, constraint element, and target property to generate composition-dependent maps such as solid-solution phase fraction or number of predicted phases. (b) Phase-fraction module for a fixed multicomponent composition, showing the predicted evolution of equilibrium phase fractions as a function of temperature. (c) Binary/ternary phase-diagram module, where users can generate conventional binary temperature–composition diagrams or isothermal ternary phase diagrams. The interface provides an interactive front end to the thermodynamic prediction workflow and allows users to download both the generated figures and the underlying numerical data.